\documentclass[fleqn,usenatbib]{mnras}

\usepackage{newtxtext,newtxmath}
\usepackage[T1]{fontenc}
\usepackage{ae,aecompl}

\usepackage{graphicx}	% Including figure files
\usepackage{amsmath}	% Advanced maths commands
\usepackage{amssymb}	% Extra maths symbols

\usepackage{enumitem,kantlipsum}

\pdfoutput=1

\title[Evolution of spatially resolved SFMS]{Evolution of spatially resolved star formation main sequence and surface density profiles in massive disc galaxies at $0\lesssim z \lesssim 1$: inside-out stellar mass buildup and quenching}

\author[Abdurro'uf \& Masayuki Akiyama]{
Abdurro'uf$^{1}$\thanks{E-mail: abdurrouf@astr.tohoku.ac.jp}
and Masayuki Akiyama$^{1}$
\\
$^{1}$Astronomical Institute, Tohoku University, Aramaki, Aoba, Sendai 980-8578, Japan
}

\date{Accepted XXX. Received YYY; in original form ZZZ}

\pubyear{2018}

\begin{document}
\label{firstpage}
\pagerange{\pageref{firstpage}--\pageref{lastpage}}
\maketitle

\begin{abstract}
We investigate a relation between surface densities of star formation rate (SFR) and stellar mass ($M_{*}$) at a $\sim 1$ kpc scale namely spatially resolved star formation main sequence (SFMS) in massive ($\log(M_{*}/M_{\odot})>10.5$) face-on disc galaxies at $0.01<z<0.02$ and $0.8<z<1.8$ and examine evolution of the relation. The spatially resolved SFMS of $z\sim 0$ galaxies is discussed in a companion paper. For $z\sim 1$ sample, we use 8 bands imaging dataset from CANDELS and 3D-HST and perform a pixel-to-pixel SED fitting to derive the spatially resolved SFR and $M_{*}$. We find a linear spatially resolved SFMS in the $z\sim 1$ galaxies that lie on the global SFMS, while a 'flattening' at high $\Sigma_{*}$ end is found in that relation for the galaxies that lie below the global SFMS. Comparison with the spatially resolved SFMS of the $z\sim 0$ galaxies shows smaller difference in the specific SFR (sSFR) at low $\Sigma_{*}$ than that at high $\Sigma_{*}$. This trend is consistent with the evolution of the sSFR$(r)$ radial profile, which shows a faster decrease in the central region than in the outskirt, agrees with the inside-out quenching scenario. We then derive an empirical model for the evolution of the $\Sigma_{*}(r)$, $\Sigma_{\rm SFR}(r)$ and sSFR$(r)$ radial profiles. Based on the empirical model, we estimate the radial profile of the quenching timescale and reproduce the observed spatially resolved SFMS at $z\sim 1$ and $z\sim 0$. 
\end{abstract}


%%%%%%%%%%%%%%%%%%%%%%%%%%%%%%%%%%%%%%%%%%%%%%%%%%

%%%%%%%%%%%%%%%%% BODY OF PAPER %%%%%%%%%%%%%%%%%%

\section{Introduction}
Thanks to an increasing number of wide-field and multi-wavelength imaging dataset, we can study scaling relations of galaxies in a wide redshift range. Investigation on properties of galaxies found that there is a linear correlation between the integrated star formation rate (SFR) and stellar mass ($M_{*}$) of star-forming galaxies namely star formation main sequence (SFMS) \citep[e.g.][]{brinchmann2004, daddi2007, elbaz2007, noeske2007, salim2007, whitaker2012}. The integrated SFMS relation suggests that SFR increases with $M_{*}$ as a power law (SFR $\propto M_{*}^{\alpha}$  with $\alpha \sim 1$) over at least two orders of magnitude in stellar mass ($\sim10^{9}-10^{11}M_{\odot}$) up to $z\sim 6$. The normalization of the integrated SFMS relation shows a factor of $\sim 2$ dex decrease from $z\sim 6$ to $z\sim 0$ \citep{speagle2014}. Tightness of the integrated SFMS relation, $1 \sigma$ scatter of only $\sim 0.3$ dex in the redshift range of $0\lesssim z \lesssim 3$ \citep[e.g.][]{whitaker2012, speagle2014, kurczynski2016}, implies an importance of a continuous internal secular process in driving the star formation activity of the majority of galaxies rather than stochastic merger process \citep{noeske2007}.

Star-forming galaxies are evolving with cosmic time maintaining their position within $\pm 0.3$ dex around the global SFMS. Once the star formation activity in a galaxy is quenched, the galaxy will move away from the global SFMS relation until it reaches the red-sequence which is populated by quiescent galaxies. Responsible mechanisms for the quenching process are still unclear. Several quenching mechanisms have been proposed. Rapid gas consumption by starburst event can make galaxies to run out their gas and in combination with the outflow driven by a stellar feedback can quench star formation in the galaxies \citep[e.g.][]{murray2005}. Furthermore, feedback from a central super massive black hole growth process [i.e. active galactic nucleus (AGN) feedback] can suppress the cold gas supply to galaxies through quasar or radio feedback mode \citep[e.g.][]{sanders1988, silk1998, springel2005, hopkins2006, hopkins2008, schawinski2006, fabian2012}. On the other hand, morphological quenching scenario proposes that once central spheroidal component (i.e. bulge) is formed, the deeper gravitational potential of the bulge can stabilize gas in the disc, and the stabilization prevents gas collapse and stop the star formation in the disc \citep[e.g.][]{martig2009, genzel2014}. The suppression of cold gas accretion into a galaxy will also happen once the growth of host dark matter halo mass reaches a certain critical mass ($\sim 10^{12}M_{\odot}$) above which newly accreted gas will be shock heated \citep[e.g.][]{birnboim2003, dekel2006}.

Investigation on the morphology and structural properties of star-forming and quiescent galaxies revealed that quiescent galaxies tend to have higher S\' ersic index ($n$) and concentration index, i.e. higher bulge fraction (\textit{B/T}, bulge-to-total mass ratio), than star-forming galaxies \citep[e.g.][]{kauffmann2003a, wuyts2011}. It is still unclear how galaxies change their morphology from disc-dominated system (low concentration index and sersic index, $n \sim 1$) to bulge-dominated system (high concentration index and sersic index, $n\gtrsim 3$).  Investigation on the radial stellar mass surface density profiles of massive galaxies at $0\lesssim z \lesssim 3$ revealed that massive galaxies establish their structures and stellar masses in a 'inside-to-outside' manner, where a bulge is already formed at $z\sim 2$ then a disc component is build subsequently \citep[e.g.][]{vandokkum2010, schreiber2011, nelson2012, nelson2016, patel2013, morishita2015, tacchella2015, tadaki2017}. Although it is suggested that galaxies change their morphologies to a bulge-dominated system during the quenching process, other investigation suggests that quiescent galaxies were born as a bulge-dominated system \citep{abramson2016}. 

As the stellar mass buildup progresses inside-out, the quenching process also happen in the similar manner. This 'inside-out quenching' process is imprinted in the positive gradient of specific SFR (sSFR) radial profile of massive galaxies at $0\lesssim z \lesssim 2$ \citep[e.g.][]{tacchella2015, tacchella2018, gonzalez2016, abdurrouf2017, belfiore2018}. It is still unclear what is a physical mechanism responsible for the inside-out quenching. Some simulation works have been done to study the physical mechanism behind the inside-out quenching. Cosmological zoom-in simulations done by \citet{zolotov2015} and \citet{tacchella2016a, tacchella2016b} suggest that galaxy may experience central gas compaction followed by a central starburst which consumes gas rapidly in the central region. If further cold gas supply into the central region is stopped due to radiative stellar feedback and/or AGN feedback, the onset of the inside-out quenching begin.

To understand how galaxy's internal star formation leads to the building up of the galaxy's stellar mass and structure and also to understand how an internal quenching process shut down the star formation in the galaxy, an analysis on the spatially resolved distributions of $M_{*}$ and SFR for a large number of galaxies in a wide redshift range is essential. Recently, investigations on sub-galactic ($\sim1$ kpc-scale) surface densities of stellar mass ($\Sigma_{*}$) and SFR ($\Sigma_{\rm SFR}$) of $z\sim 0$ and $z\sim 1$ galaxies revealed that there is a nearly linear relation between $\Sigma_{*}$ and $\Sigma_{\rm SFR}$ in a similar form as found in the integrated scaling relation, namely spatially resolved SFMS relation (for $z\sim 1$: \citet{wuyts2013} and \citet{magdis2016}, while for $z\sim 0$: e.g. \citet{canodiaz2016}, \citet{maragkoudakis2017}, \citet{abdurrouf2017}, \citet{hsieh2017}, \citet{medling2018}, and \citet{liu2018}). Previous research papers reported the spatially resolved SFMS relations with various slopes ($\sim 0.7-1.0$) and zero points. This discrepancy is possibly caused by different methods used in each research, especially on the SFR indicator, i.e. method to derive SFR \citep{speagle2014}. 

Understanding the spatially resolved SFMS relation and its evolution with cosmic time is very important to study the origin of the global SFMS relation, because the sub-galactic relation can be a more fundamental relation from which the global relation is originated. \citet{abdurrouf2017} studied the spatially resolved SFMS relation in the local ($0.01<z<0.02$) massive ($M_{*}>10^{10.5}M_{\odot}$) disc galaxies using seven bands (FUV, NUV, $u$, $g$, $r$, $i$ and $z$) imaging data from \textit{Galaxy Evolution Explorer} (\textit{GALEX}) and Sloan Digital Sky Survey (SDSS). In that research, we derived the spatially resolved SFR and stellar mass of a galaxy by using a method so-called pixel-to-pixel spectral energy distribution (SED) fitting which fits the spatially resolved SED of a galaxy with a set of model photometric SEDs using a Bayesian statistics approach. The reason for choosing the method is that the same method is applicable to a large number of galaxies even at high redshifts, thanks to the high spatial resolution of the near-infrared (NIR) images taken by the \textit{Hubble Space Telescope} (\textit{HST}).

\citet{abdurrouf2017} found that the spatially resolved SFMS in the local massive disc galaxies show that $\Sigma_{\rm SFR}$ increases linearly with $\Sigma_{*}$ at low $\Sigma_{*} (\lesssim 10^{7.5}M_{\odot}\text{kpc}^{-2})$ range, while flattened at high $\Sigma_{*} (\gtrsim 10^{7.5}M_{\odot}\text{kpc}^{-2})$ range. Investigation on the spatially resolved SFMS relation in the galaxies above $+0.3$ dex (hereafter, z0-$\Delta$MS1), between $-0.3$ and $+0.3$ dex (hereafter, z0-$\Delta$MS2) and below $-0.3$ dex (hereafter z0-$\Delta$MS3) of the global SFMS relation, found a tight spatially resolved SFMS relation in the z0-$\Delta$MS1 and z0-$\Delta$MS2 galaxies, while the relation seems to be broken in the z0-$\Delta$MS3 galaxies. The normalization in the global SFMS in each group is preserved in the spatially resolved SFMS, in the sense that the spatially resolved SFMS of z0-$\Delta$MS1 galaxies has higher normalization than the spatially resolved SFMS of z0-$\Delta$MS2 galaxies.   

In the current work, we extend our previous study of the spatially resolved SFMS to massive disc galaxies at $0.8<z<1.8$ using the similar pixel-to-pixel SED fitting method applied to the 8 bands (F435W, F606W, F775W, F814W, F850LP, F125W, F140W and F160W) imaging data from the Cosmic Assembly Near-infrared Deep Extragalactic Legacy Survey \citep[CANDELS;][]{grogin2011, koekemoer2011} and 3D-HST \citep{brammer2012} projects. Similar rest-frame wavelength coverage (FUV-NIR) and spatial resolution ($\sim 1$ kpc) of the imaging data used in this work to those in the previous work allows a consistent comparison and could resolve the problem caused by the different method in studying the spatially resolved SFMS at different redshifts. Furthermore, we also discuss the evolution of the $\Sigma_{*}$, $\Sigma_{\rm SFR}$ and sSFR radial profiles.

The structure of this paper is as follows. In Section~\ref{sec:data_sample}, we explain the sample. Section~\ref{sec:method} presents our methodology, pixel-to-pixel SED fitting. Results and discussions are presented in Section~\ref{sec:results} and \ref{sec:discussion}, respectively. The cosmological parameters of $\Omega_{m}=0.3$, $\Omega_{\Lambda}=0.7$ and $H_{0}=70 \text{km} \text{s}^{-1} \text{Mpc}^{-1}$ are used throughout this paper. We use $M_{*}$ to represent the total stellar mass of a galaxy, while $m_{*}$ is used to represent the stellar mass within a sub-galactic region. Terms global is used to indicate a galaxy-scale quantity, while term sub-galactic is used to represent $\sim 1$ kpc scale quantity within a galaxy.    

\section{Data sample}
\label{sec:data_sample}
To examine the relation between $\Sigma_{*}$ and $\Sigma_{\rm SFR}$ of galaxies at $z\sim 1$ with the same resolution of 1-2 kpc as those of local galaxies in the companion paper \citep{abdurrouf2017}, we use eight bands (F435W, F606W, F775W, F814W, F850LP, F125W, F140W and F160W) imaging data from CANDELS \citep[]{grogin2011, koekemoer2011} and 3D-HST \citep{brammer2012} which cover $\sim 4000${\AA}$-16000${\AA}. The eight bands at $z\sim 1$ have similar rest-frame wavelength coverage to the seven bands (FUV, NUV, $u$, $g$, $r$, $i$ and $z$) of \textit{GALEX} and SDSS imaging data for galaxies at $z\sim 0$. Thanks to the wide wavelength coverage, degeneracy between age and dust extinction (inherent in the stellar population synthesis models) can be broken. The dust extinction can be constrained by the rest-frame FUV$-$NUV colour (observed F435W$-$F606W colour at $z\sim 1$), while age can be constrained by the rest-frame $u-g$ colour (observed F775W$-$F850LP colour at $z\sim 1$).

We select sample galaxies located in the GOODS-S field from the 3D-HST catalog \citep[]{skelton2014, brammer2012} based on $M_{*}$ and redshift. In the catalog, the $M_{*}$ is calculated through $0.3\mu m$ to $8\mu m$ SED modeling using the FAST code \citep{kriek2009} and redshift is determined using three methods: (1) photometric redshifts using the $0.3\mu m$ to $8\mu m$ SED fitting with EAZY code \citep{brammer2008}, (2) two-dimensional grism spectroscopy by the 3D-HST and (3) ground-based spectroscopy. For SFR, we do not use SFR derived by the FAST code, instead we use the SFR calculated following \citet{whitaker2014}, which uses the combination of rest-frame UV and IR luminosities. We applied following criteria to select the sample galaxies: (1) redshift range of $0.8<z<1.8$, (2) stellar mass higher than $10^{10.5}M_{\odot}$, (3) observed in the eight bands, (4) face-on configuration with ellipticity less than $0.6$ or $b/a > 0.4$ and (5) late-type (disc) morphology with S\' ersic index ($n$) less than $2.6$.

The redshift range is determined to achieve resolution of $\sim 1$ kpc with F160W image, which has largest full width at half-maximum (FWHM) with $0.19$ arcsec among the eight bands. In the redshift range, the eight band coverage samples the rest-frame SED in the FUV to near-infrared (NIR) wavelength. We apply the same mass limit of $10^{10.5}M_{\odot}$ as in \citet{abdurrouf2017}. We select face-on galaxies to minimize the effect of dust extinction. The ellipticities of the galaxies are calculated by averaging the F125W-band elliptical isophotes outside of an effective radius, as described in the construction of the radial profiles (see Section~\ref{sec:radial_profiles}). The S\' ersic index is calculated based on S\' ersic profile fitting to the one-dimensional stellar mass surface density radial profile ($\Sigma_{*}(r)$) using the maximum likelihood method. The calculation of the $\Sigma_{*}(r)$ and S\' ersic index are explained in Section~\ref{sec:radial_profiles}. In addition to the five selection criteria described above, we only select galaxies which have more than four bins of pixels, where each bin has a signal-to-noise (S/N) ratio of more than $10$ in all of the eight bands (see Section 3.2 of \citet{abdurrouf2017} for the description on the binning method).

\begin{figure}
\centering
\includegraphics[width=0.5\textwidth]{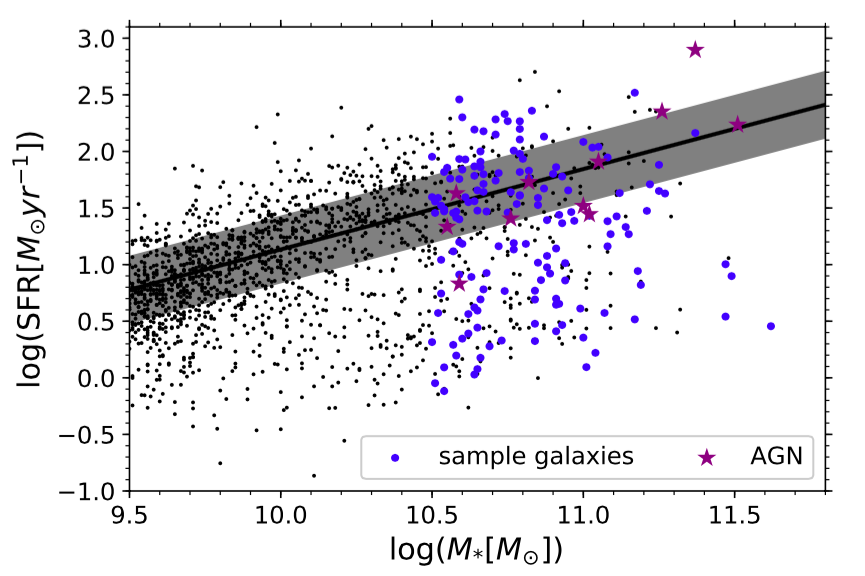}
\includegraphics[width=0.5\textwidth]{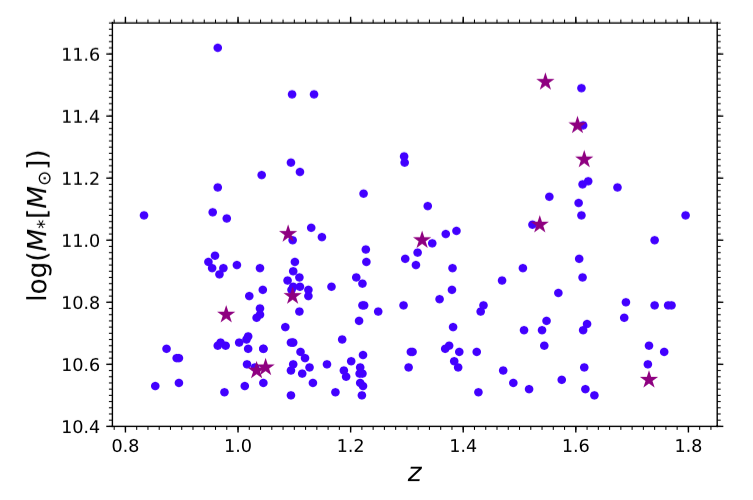}
\caption{Top panel: SFR versus $M_{*}$ of galaxies more massive than $\log(M_{*}/M_{\odot})=9.5$ at $0.8<z<1.8$ located in the GOODS-S region. Blue circles represent $152$ sample galaxies used in this work, while the purple star symbols represent AGN-host galaxies. The black line represents the global SFMS relation of \citet{speagle2014} calculated at the median redshift of the sample galaxies, $z=1.217$, and the gray shaded area around it represents the $\pm 0.3$ dex scatter. Bottom panel: redshifts versus $M_{*}$ of the sample galaxies. \label{fig:galaxies_sample}}
\end{figure}

For galaxies with the photometric redshift, we check the reliability of the redshift estimation by fitting the integrated SEDs of the galaxies in the F435W to F160W bands with the model SEDs, which are calculated at the redshifts of the galaxies. We use the maximum likelihood method to get the best-fitting model SED. We find eight galaxies with a strange SED which results in a very large $\chi^{2}$ and lead to an unreliable redshift estimate, while the other galaxies have small $\chi^{2}$, which indicates the reliability of their photometric redshifts. We then exclude the eight galaxies from the sample. Finally, we cross-match the remaining 163 galaxies with the \textit{Chandra} $7$ Ms sources catalog \citep[]{luo2017, yang2017} to remove galaxies with a luminous AGN activity. The \textit{Chandra} catalog contains X-ray sources from the $\sim 7$ Ms exposure in the \textit{Chandra} Deep Field-South (CDF-S), which covers GOODS-S field. We find 11 galaxies that have a luminous X-ray AGN activity ($L_{2-10\text{keV}}>10^{43}\text{erg }\text{s}^{-1}$) among the sample. We then exclude those galaxies from the sample to avoid contamination by the contribution from the non-stellar AGN component to the broad band photometry. Finally, $152$ galaxies are selected for further analysis. Top panel of Fig.~\ref{fig:galaxies_sample} shows the $M_{*}$ and SFR of the $152$ sample galaxies (blue circles) along with the distribution of entire galaxies more massive than $10^{9.5}M_{\odot}$ at $0.8<z<1.8$ in the GOODS-S field (small black circles). The black line indicates global SFMS relation of \citet{speagle2014} calculated at the median redshift of the sample, $z=1.217$, while the gray shaded area represents $\pm 0.3$ dex scatter around the global SFMS relation. The purple stars represent AGN-host galaxies. Bottom panel of Fig.~\ref{fig:galaxies_sample} shows redshift versus $M_{*}$ of the sample galaxies. The figure shows that redshifts of the sample galaxies are spread uniformly within the redshift range.   

The eight band mosaic images from the 3D-HST\ \footnote{\url{http://3dhst.research.yale.edu/Data.php}} are registered to the same sampling of $0.06$ $\text{arcsec }\text{pixel}^{-1}$ and PSF-matched to the F160W image. Background of the mosaic images are subtracted. The FWHM corresponds to the physical scale of $\sim 1.4-1.6$ kpc at $0.8<z<1.8$. $5\sigma$ limiting magnitudes of the F435W, F606W, F775W, F814W, F850LP, F125W, F140W and F160W are $27.3$, $27.4$, $26.9$, $27.2$, $26.5$, $26.1$, $25.6$ and $26.4$ mag within $0.7$ arcsec diameter, respectively \citep{skelton2014}.

\section{Methodology} \label{sec:method}
In order to derive spatially resolved stellar population properties, especially SFR and $M_{*}$, we use a method namely pixel-to-pixel SED fitting, which is the same method as we used in \citet{abdurrouf2017}. We fit spatially resolved SED of each bin with a set of model SEDs using a Bayesian statistics approach. The method can be divided into three main steps: (1) Image registration, PSF matching, and pixel binning to get photometric SED of each bin of a galaxy, (2) construction of a library of model photometric SEDs, and (3) fitting the SED of each bin with the set of the model SEDs, as described in detail in \citet{abdurrouf2017}.

We do not need image registration and PSF matching because eight bands imaging data provided by the 3D-HST have been registered and PSF-matched as described previously, so the first step is to define an area of a galaxy. To define the area of a galaxy, we firstly generate a segmentation map for the mosaic image in each band using \texttt{SExtractor} \citep{bertin1996} with a detection threshold of above 1.5 times larger than the rms scatter outside of the galaxy, then using the position of a specific galaxy from the 3D-HST catalog, we find the segmentation map around the galaxy. In the \texttt{SExtractor} segmentation map, each object is indicated with a different value, which correspond to the id number of the object in the generated catalog. By reading the pixel value of the galaxy's central pixel and looking for other pixels which have the same value, we can obtain pixels associated with the galaxy. Some outlier pixels which are not connected with the main area of the galaxy are sometimes included in the area of the galaxy, in such case, we exclude those pixels which have no connection to the central pixel of the galaxy. The segmentation maps of the galaxy in eight bands are then merged to define the area of the galaxy. 

\begin{figure}
\centering
\includegraphics[width=0.5\textwidth]{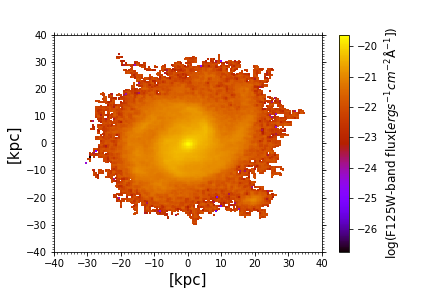}
\includegraphics[width=0.5\textwidth]{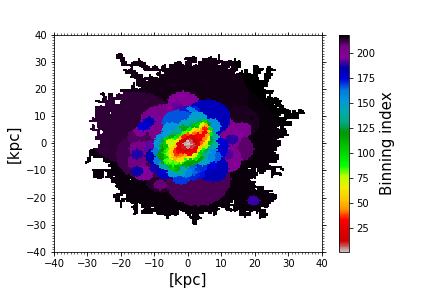}
\caption{Top panel: F125W band image of a galaxy \texttt{GS\_19186} in the sample which located at RA$=53^{\circ}.120750$, DEC$=-27^{\circ}.818984$ and $z=1.0940$. Bottom panel: pixel binning result of the galaxy. Color coding represents index of bin. North is to the top and east is to the left. \label{fig:binningmap}}
\end{figure}

The second step is converting a pixel value in a unit of $\text{count }\text{s}^{-1}$ to the flux in $\text{erg }\text{s}^{-1} \text{cm}^{-2}${\AA}$^{-1}$ and then pixel binning to increase the S/N ratio. The pixel value to flux conversion is done by multiplying the pixel value with a conversion factor given in the PHOTFLAM header keyword. The pixel binning is done by considering not only an S/N threshold to be reached by combining the pixels, but also similarity of SED shape (tested through a $\chi^{2}$ calculation) among the pixels which will be binned. The pixel binning is done by first, looking for the brightest pixel in the F125W band, then check each neighboring pixel located within a circular annulus centered at the brightest pixel, for the similarity of its SED shape to that SED of the brightest pixel (with $\chi^{2}$ below a certain limit) and include the pixel into the bin if its SED shape is similar. Radius of the circular annulus is then increased by $2$ pixels and the same procedure is done to add up more pixels until the total S/N of the bin reaches the S/N threshold. Next bin is made by looking for the brightest pixel in the F125W band that was not included in the previous binning and then do the same steps as above. The above procedure is applied until no bin can be made with the remaining pixels. Finally, all the remaining pixels are binned together into one bin. Similar as in \citet{abdurrouf2017}, we set S/N threshold of 10 in all eight bands and $\chi^{2}$ limit of 30. Fig.~\ref{fig:binningmap} shows (top panel) the F125W band image and (bottom panel) the binning result of a galaxy \texttt{GS\_19186}, which is located at RA$=53^{\circ}.120750$, DEC$=-27^{\circ}.818984$ and $z=1.0940$. 

The next step is constructing a library of model SEDs. A library of $300,000$ model photometric SEDs with a random set of parameters [$\tau$, $t$, $E(B-V)$, and $Z$] is generated by interpolating the $286,000$ parent model SEDs in a grid of those parameters. We use GALAXEV stellar population synthesis model \citep{bruzual2003} with \citet{chabrier2003} initial mass function (IMF) and exponentially declining star formation history of $\text{SFR}(t)\propto e^{-t/\tau}$. $\tau$, $t$, $E(B-V)$, and $Z$ represent SFR decaying timescale, age of the stellar population, color excess of dust attenuation, and metallicity of the stellar population, respectively. We multiply parent model spectra with the eight filter transmission curves of CANDELS and 3D-HST then integrate to get model fluxes in the $8$ bands. To apply effect of dust extinction, we use \citet{calzetti2000} dust extinction law. The random set of parameters have ranges of: $\tau[0.1:10\text{ Gyr}]$, $t[0.25:6.6\text{ Gyr}]$, $E(B-V)[0:0.6\text{ mag}]$ and $Z[0.004:0.05]$. Those parameter ranges are the same as those used in \citet{abdurrouf2017}, except for the age range for which the age of the universe at $z=0.8$ is used as an upper limit. As in the previous work, the interpolation to estimate $8$ band fluxes and stellar masses for a random parameter set is done in two steps, first interpolation in a three-dimensional space [$E(B-V)$, $t$ and $\tau$] using a tricubic interpolation for each metallicity ($Z$), then in one-dimensional space of $Z$ with a cubic spline interpolation.   

After constructing the spatially resolved SED of a galaxy and generating the library of the model SEDs, the final step is fitting the observed SED of each bin with the library of the model SEDs to get $m_{*}$ and SFR of the bin. The fitting is done using a Bayesian statistics approach. In this approach, probability distribution functions (PDFs) of the $m_{*}$ and SFR are constructed by compounding probabilities of the model SEDs, then posterior means of the $m_{*}$ and SFR are calculated. We evaluate probability of a model based on its $\chi^{2}$ in a form of Student's t distribution with degree of freedom, $\nu$, of 3, instead of a Gaussian form. It has been verified that, this new model weighting scheme gives a consistent estimate of SFR and sSFR with those estimated from $24\mu$m flux (see appendix A of \citet{abdurrouf2017}). Uncertainties of the SFR and $m_{*}$ are estimated by calculating standard deviation of the PDFs of SFR and $m_{*}$. Once the $m_{*}$ and SFR of a bin are obtained, those values are then divided into the pixels that belong to the bin by assuming that the $m_{*}$ and SFR of a pixel are proportional to the pixel's fluxes in F160W and F435W bands, respectively.  

\begin{figure*}
\centering
\includegraphics[width=0.9\textwidth]{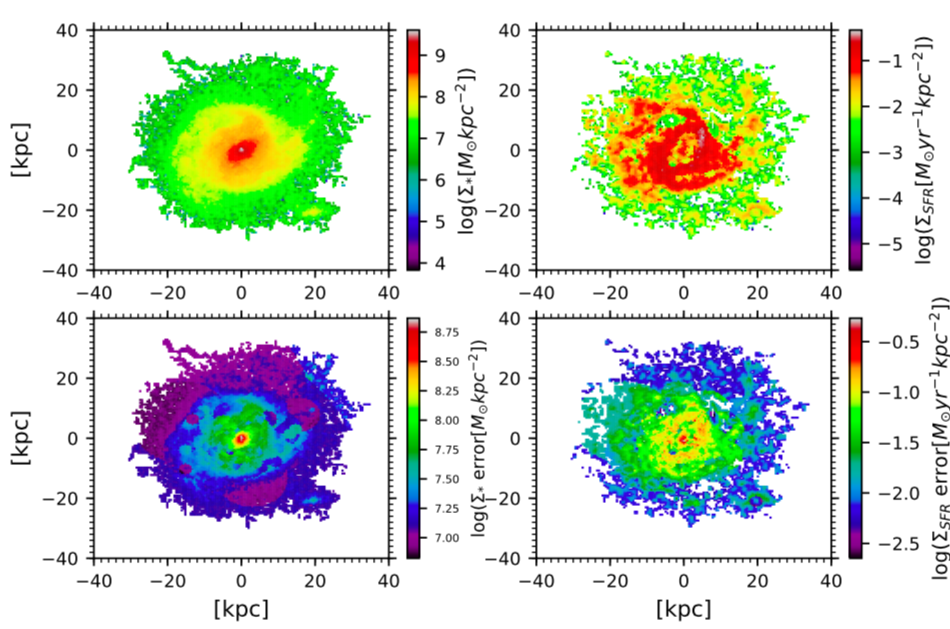}
\caption{An example of pixel-to-pixel SED fitting result for a galaxy \texttt{GS\_19186} (RA$=53^{\circ}.120750$, DEC$=-27^{\circ}.818984$ and $z=1.0940$). Top left panel: stellar mass surface density ($\Sigma_{*}$) map. Top right panel: SFR surface density ($\Sigma_{\rm SFR}$) map. Bottom left panel: $\Sigma_{*}$ uncertainty map. Bottom right panel: $\Sigma_{\rm SFR}$ uncertainty map. Up and left directions correspond to the north and east, respectively.  \label{fig:fitting_results}}
\end{figure*}

Fig.~\ref{fig:fitting_results} shows an example of the pixel-to-pixel SED fitting result for a galaxy \texttt{GS\_19186} in the sample (whose pixel binning result is shown in Fig.~\ref{fig:binningmap}). The $\Sigma_{\rm SFR}$ map roughly traces spiral arms which are associated with high star formation activity, while the $\Sigma_{*}$ map shows smoother distribution. Pixels with negative value in the $\Sigma_{\rm SFR}$($\Sigma_{*}$) due to a negative value in the F435W(F160W) flux caused by noise fluctuation, are not shown in the plot with the logarithmic scale. Those pixels with negative values are included in the later analysis e.g. calculation of the integrated SFR and $M_{*}$ and radial profiles of $\Sigma_{\rm SFR}(r)$ and $\Sigma_{*}(r)$.              

\begin{figure}
\centering
\includegraphics[width=0.5\textwidth]{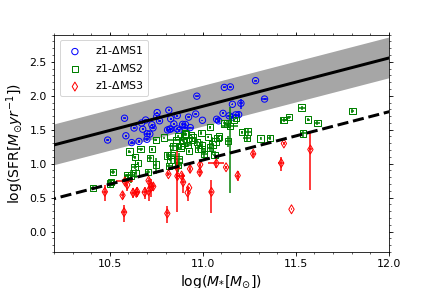}
\caption{Integrated SFR versus $M_{*}$ of the sample galaxies. The solid line represents the global SFMS relation of \citet{speagle2014} calculated at median redshift of the sample ($z=1.217$) and the dashed line represents SFMS$-0.8$ dex. Grey-shaded region corresponds to $\pm 0.3$ dex from the global SFMS. The blue circles (located within $\pm 0.3$ dex about the global SFMS), green squares (located between $-0.3$ and $-0.8$ dex from the global SFMS) and red diamonds (located below $-0.8$ dex from the global SFMS) represent z1-$\Delta$MS1, z1-$\Delta$MS2 and z1-$\Delta$MS3 sub-samples, respectively. \label{fig:integrated_SFMS}}
\end{figure}

\begin{figure}
\centering
\includegraphics[width=0.5\textwidth]{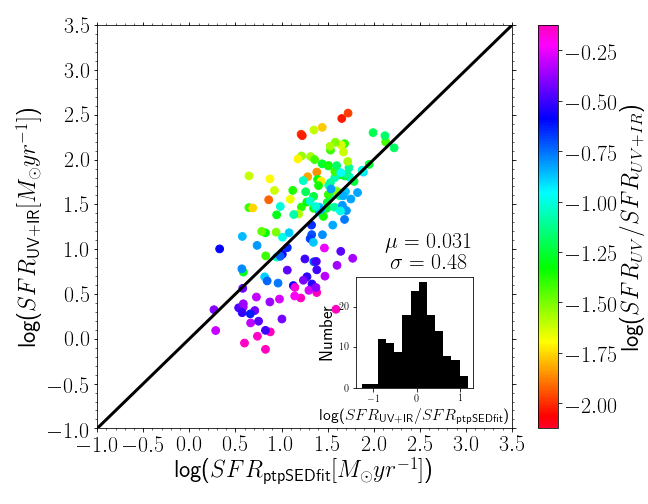}
\caption{Comparison between the integrated SFR estimated using the pixel-to-pixel SED fitting method ($\text{SFR}_{\rm ptpSEDfit}$) and that from the 3D-HST catalog ($\text{SFR}_{\rm UV+IR}$). The black line shows proportionality line. The histogram shows the distribution of $\log(\text{SFR}_{\rm UV+IR}/\text{SFR}_{\text{ptpSEDfit}})$ which has mean value ($\mu$) of $0.031$ and standard deviation ($\sigma$) of $0.48$ dex. The color-coding represents the ratio of $\log(\text{SFR}_{\rm UV}/\text{SFR}_{\rm UV+IR})$, which is expected to be inversely proportional to the amount of dust extinction. \label{fig:compare_SFR_ptpSEDfit_3DHST}}
\end{figure}

%%%%%%%%% make change here:
Fig.~\ref{fig:integrated_SFMS} shows the integrated SFR versus $M_{*}$ of the sample galaxies. The integrated SFR and $M_{*}$ of a galaxy are derived by summing up the SFR and $m_{*}$ of all pixels that belong to the galaxy. Distributions of the sample galaxies on the SFR versus $M_{*}$ plane derived with our method (which is shown in Fig.~\ref{fig:integrated_SFMS}) is considerably different compared to that with the 3D-HST catalog (which is shown in Fig.~\ref{fig:galaxies_sample}). This discrepancy is caused by the discrepancy in the estimation of both of the SFR and $M_{*}$. Fig.~\ref{fig:compare_SFR_ptpSEDfit_3DHST} shows comparison between the integrated SFR derived using our pixel-to-pixel SED fitting method ($\text{SFR}_{\rm ptpSEDfit}$, which is the sum of the SFR of galaxy's pixels) and that from the 3D-HST catalog ($\text{SFR}_{\rm UV+IR}$). It is shown by the figure that the $\text{SFR}_{\rm ptpSEDfit}$ is broadly consistent with the $\text{SFR}_{\rm UV+IR}$. The Histogram shows the distribution of the $\log(\text{SFR}_{\rm UV+IR}/\text{SFR}_{\text{ptpSEDfit}})$, which has a mean value ($\mu$) of $0.031$ and a standard deviation ($\sigma$) of $0.48$ dex. The color-coding represents the ratio of $\log(\text{SFR}_{\rm UV}/\text{SFR}_{\rm UV+IR})$ which is expected to be inversely proportional to the amount of dust extinction. It is shown by the figure that there is a systematic dependence on the amount of dust extinction. It is suggested that the estimated summed SFR from the pixel-to-pixel SED fitting is systematically smaller for galaxies with large dust extinction. The large offset is only observed among a few galaxies in the sample and we expect their effects on the analysis of the statistical sample can be minor. The issues of the discrepancies in the SFR and $M_{*}$ are further discussed in appendix A. The $M_{*}$ estimated using our method systematically higher than the $M_{*}$ taken from the 3D-HST catalog (see lower panel in Fig.~\ref{SFR_SM_3DHST_vs_ptpSEDfit}). 

In later analysis, we will discuss the difference between spatially resolved SFMS relations of galaxies as a function of their distances from the global SFMS relation in the SFR versus $M_{*}$ plane. As we used the global SFMS relation by \citet{speagle2014} to classify galaxies based on their distances from the global SFMS in \citet{abdurrouf2017}, here we also use the same global SFMS relation. The solid line in Fig.~\ref{fig:integrated_SFMS} represents the global SFMS relation calculated at median redshift of the sample, $z=1.217$. The grey-shaded area represents $\pm 0.3$ dex around the global SFMS relation. Galaxies that are located within $\pm 0.3$ dex, between $-0.3$ and $-0.8$ dex, and below $-0.8$ dex from the global SFMS are called z1-$\Delta$MS1 (blue circle), z1-$\Delta$MS2 (green square), and z1-$\Delta$MS3 (red diamond), respectively. The above sSFR groups are selected such that majority of the z1-$\Delta$MS1 and z1-$\Delta$MS3 are star-forming and quiescent galaxies, respectively. The upper limit for defining the z1-$\Delta$MS3 is chosen such that majority of galaxies below the upper limit are quiescent galaxies, by verifying it with the $UVJ$ diagram. Positions of the sample galaxies on the $UVJ$ diagram and verification that majority of the z1-$\Delta$MS1 and z1-$\Delta$MS3 are star-forming and quiescent galaxies, respectively, are described in appendix B. Median values of $\log(\text{sSFR}[yr^{-1}])$ (number of galaxies) of the z1-$\Delta$MS1, z1-$\Delta$MS2, and z1-$\Delta$MS3 sub-samples are $-9.28$ (47), $-9.70$ (72), and $-10.07$ (33), respectively.

\section{Results} \label{sec:results}
\subsection{Spatially resolved star formation main sequence in massive disc galaxies at \texorpdfstring{$z\sim 1$}{Lg}} \label{sec:spatially_resolved_SFMS}

\begin{figure}
\centering
\includegraphics[width=0.5\textwidth]{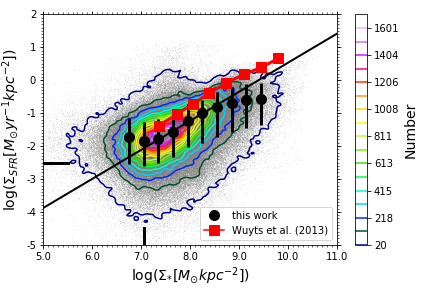}
\caption{$\Sigma_{\rm SFR}$ versus $\Sigma_{*}$ of all 597651 pixels of the 152 sample galaxies shown with contours colour-coded by number of pixels in each $0.1\times 0.1$ dex bin. The vertical (horizontal) lines at bottom (left) axes are limiting values of the $\Sigma_{*}$($\Sigma_{\rm SFR}$) which are median values of the $\Sigma_{*}$($\Sigma_{\rm SFR}$) for pixels located in outskirt (the outermost $8$ kpc elliptical annulus) of the sample galaxies. The black circles with error bars show spatially resolved SFMS relation obtained by taking mode of the $\Sigma_{\rm SFR}$ distribution for each $\Sigma_{*}$ bin with $0.3$ dex width. The black line represents linear function fitting to the five mode values with $\Sigma_{*}\lesssim 10^{8.5}M_{\odot}\text{kpc}^{-2}$, excluding two mode values in the lowest $\Sigma_{*}$. The red line with squares shows the spatially resolved SFMS relation of \citet{wuyts2013}. \label{fig:resolved_SFMS}}
\end{figure}

To examine the relation between $\Sigma_{*}$ and $\Sigma_{\rm SFR}$ at $\sim 1$ kpc scale in the $z\sim 1$ massive disc galaxies, the $\Sigma_{*}$ and $\Sigma_{\rm SFR}$ of all 597651 pixels of the sample galaxies are plotted in Fig.~\ref{fig:resolved_SFMS}. In the figure, the contours are colour-coded with the number of pixels in each $0.1 \times 0.1$ dex bin. The vertical (horizontal) lines at the bottom (left) axes are the median values of $\Sigma_{*}$($\Sigma_{\rm SFR}$) for pixels located in the outskirt (the outermost $8$ kpc elliptical annulus) of the sample galaxies and they represent the limiting values for those quantities considering the low S/N of the outskirt pixels (S/N$\sim 0.5$ per pixel). The contours with high number density imply a tight relation between $\Sigma_{*}$ and $\Sigma_{\rm SFR}$. The black circles with error bars over-plotted on the contours show the mode of $\Sigma_{\rm SFR}$ distribution for each $\Sigma_{*}$ bin with $0.3$ dex width. Error bars represent the standard deviation from the mode, and calculated separately above and below the mode value. As shown by the mode values, the relation between $\Sigma_{*}$ and $\Sigma_{\rm SFR}$ is linear at low $\Sigma_{*}$ ($\lesssim 10^{8.5}M_{\odot}\text{kpc}^{-2}$) and flattened at high $\Sigma_{*}$ end ($\gtrsim 10^{8.5}M_{\odot}\text{kpc}^{-2}$). 

Fitting the linear part of the mode values (consist of five mode values with $\Sigma_{*} \lesssim 10^{8.5}M_{\odot}\text{kpc}^{-2}$ and excluding the two lowest $\Sigma_{*}$ points, which are affected by the limiting value of $\Sigma_{*}$) with a linear relation with a form of
\begin{equation}
\log \Sigma_{\rm SFR} = \alpha \log \Sigma_{*} + \beta
\end{equation}
using a least-square fitting method resulted in the best-fitting relation with the slope ($\alpha$) of $0.88$ and zero-point ($\beta$) of $-8.31$, which is shown by the black line. The red squares show the spatially resolved SFMS relation of massive ($\log(M_{*}/M_{\odot})>10$) star-forming galaxies at $0.7<z<1.5$ reported by \citet{wuyts2013}, which was derived from the median of $\Sigma_{\rm SFR}$ distribution in each $\Sigma_{*}$ bin. They also reported the flattening tendency of the relation at high $\Sigma_{*}$ region, although not as clear as the flattening trend obtained in this work. The systematically lower spatially resolved SFMS relation found in this work compared to that reported by \citet{wuyts2013} is in part caused by the different sample selection; massive star-forming galaxies were used in \citet{wuyts2013}, while in this work, we include not only massive star-forming galaxies, but also green-valley and quiescent galaxies which have lower $\Sigma_{\rm SFR}$ for a fixed $\Sigma_{*}$ at high $\Sigma_{*}$ region as will be discussed later.     

\begin{figure*}
\centering
\includegraphics[width=0.95\textwidth]{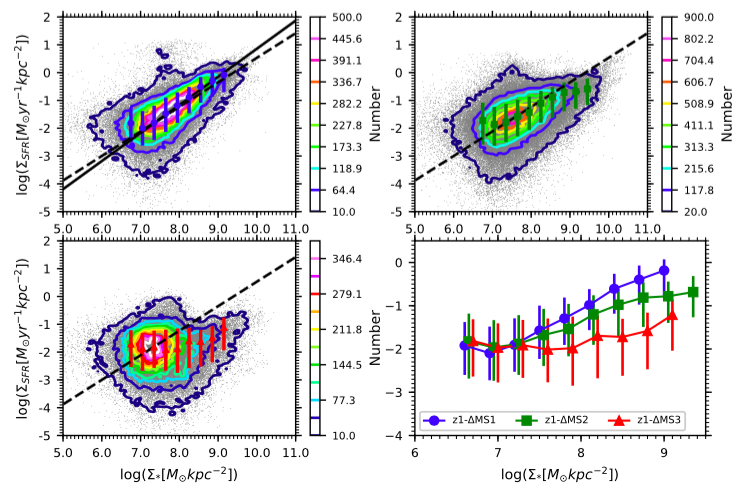}
\caption{Spatially resolved SFMS relations of the galaxies as a function of their distance from the global SFMS. Spatially resolved SFMS relation for galaxies located within $\pm 0.3$ dex from the global SFMS (z1-$\Delta$MS1; top left panel), between $-0.3$ dex and $-0.8$ dex from the global SFMS (z1-$\Delta$MS2;top right panel) and below $-0.8$ dex from the global SFMS (z1-$\Delta$MS3;bottom left panel). The spatially resolved SFMS relations of z1-$\Delta$MS1, z1-$\Delta$MS2, and z1-$\Delta$MS3 are shown with blue circles, green squares, and red triangles, respectively. The solid line in the top left panel represents the linear function fitting result to the eight mode values (excluding one with the lowest $\Sigma_{*}$) and the dashed lines in the three panels are the same as the solid line in Fig.~\ref{fig:resolved_SFMS}. Bottom right panel: comparison between the three spatially resolved SFMS relations from the previous three panels. For clarity, blue circles are shifted by 0.05 dex to the left, while red triangles are shifted by 0.05 dex to the right from their actual positions. \label{fig:resolved_SFMS_comb}}
\end{figure*}

Next, we investigate the spatially resolved SFMS relation as a function of the distance from the global SFMS. Fig.~\ref{fig:resolved_SFMS_comb} shows the spatially resolved SFMS relation in the z1-$\Delta$MS1 galaxies (top left, consists of 160210 pixels), z1-$\Delta$MS2 galaxies (top right, consists of 286721 pixels), z1-$\Delta$MS3 galaxies (bottom left, consists of 150720 pixels) and the compilation of those three relations (bottom right). The spatially resolved SFMS relations of the z1-$\Delta$MS1, z1-$\Delta$MS2, and z1-$\Delta$MS3 are shown with blue circles, green squares, and red triangles, respectively. The spatially resolved SFMS of the z1-$\Delta$MS1 galaxies shows linear increasing trend in the entire $\Sigma_{*}$ range, without flattening trend at high $\Sigma_{*}$ range as found in the spatially resolved SFMS for all galaxies (Fig.~\ref{fig:resolved_SFMS}). The flattening at high $\Sigma_{*}$ appears in the spatially resolved SFMS of the z1-$\Delta$MS2 galaxies and the flattening is more enhanced in the spatially resolved SFMS of the z1-$\Delta$MS3. The solid line in the top left panel represents the result of a linear function fitting to the eight mode values (excluding one with the lowest $\Sigma_{*}$, which is affected by the $\Sigma_{*}$ limit), which has slope of $1.01$ and zero-point of $-9.24$. The dashed lines in the three panels are the same as the solid line in the Fig.~\ref{fig:resolved_SFMS}.

Comparison between the three spatially resolved SFMS relations (bottom right panel) shows similar value of $\Sigma_{\rm SFR}$ at the low $\Sigma_{*}$ region, while there is a large difference in $\Sigma_{\rm SFR}$ at the high $\Sigma_{*}$ region. Most of the pixels associated with high $\Sigma_{*}$ are located in the central region, while the pixels associated with low $\Sigma_{*}$ are located in the disc region. The linear increasing trend of the spatially resolved SFMS of z1-$\Delta$MS1 galaxies indicates the ongoing star formation activity in the central region as well as in the outskirt, while flattening at high $\Sigma_{*}$ region in the spatially resolved SFMS relations of the other groups indicates that a quenching mechanism is ongoing in the central region.

\subsection{Radial profiles of \texorpdfstring{$\Sigma_{*}(r)$}{Lg}, \texorpdfstring{$\Sigma_{\rm SFR}(r)$}{Lg} and sSFR\texorpdfstring{$(r)$}{Lg} at \texorpdfstring{$z\sim 1$}{Lg}}
\label{sec:radial_profiles}

Increasing $\Sigma_{*}$ along the x-axis of the spatially resolved SFMS plot (Fig.~\ref{fig:resolved_SFMS} and Fig.~\ref{fig:resolved_SFMS_comb}) roughly corresponds to a decreasing radius toward the central region of the galaxies because the radial profile of $\Sigma_{*}(r)$ is always decreasing from the central region to the outskirt. Therefore, the spatially resolved SFMS might be correlated with the radial profiles of $\Sigma_{*}(r)$, $\Sigma_{\rm SFR}(r)$ and sSFR$(r)$. Here, we derive those radial profiles to study how they correlate with the spatially resolved SFMS and also study how those radial profiles change with the distance of the galaxy from the global SFMS.

First, $\Sigma_{\rm SFR}(r)$ and $\Sigma_{*}(r)$ profiles are constructed by averaging the $\Sigma_{\rm SFR}$ and $\Sigma_{*}$ of pixels in each elliptical annulus of radius $r$. Then sSFR$(r)$ profile is obtained by dividing $\Sigma_{\rm SFR}(r)$ with $\Sigma_{*}(r)$. An ellipsoids are determined as follows. First, fitting the elliptical isophotes to the F125W-band image of a galaxy using an \texttt{ellipse} command in \texttt{IRAF}. Then an average ellipticity and position angle are derived based on the ellipsoids outside of a half-mass radius of the galaxy, which is defined as the length of a semi-major axis that encloses half of the total $M_{*}$. The half-mass radius is calculated based on the ellipsoids with ellipticity and position angle that are determined by averaging the ellipticity and position angle of the entire radius. The radial profile is sampled with a $2$ kpc step.

\begin{figure*}
\centering
\includegraphics[width=1\textwidth]{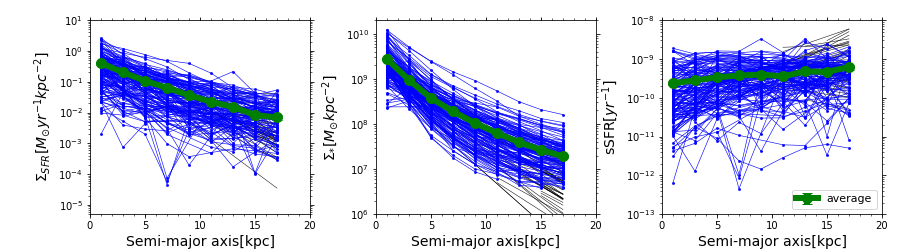}
\caption{Radial profiles of $\Sigma_{\rm SFR}(r)$ (left panel), $\Sigma_{*}(r)$ (middle panel) and sSFR$(r)$ (right panel) of all 152 sample galaxies are shown by blue squares with lines. The radial profiles are cutted up to semi-major axis of $17$ kpc. The black lines are the extrapolated part of the radial profiles which are calculated by an exponential function fitting to the region with semi-major axis larger than $3$ kpc using a least-square fitting method. The average radial profiles are shown by green circles. Errorbars for the average radial profiles are calculated by standard error of mean. \label{fig:radial_profile_combine}}
\end{figure*}

Fig.~\ref{fig:radial_profile_combine} shows the radial profiles of $\Sigma_{\rm SFR}(r)$ (left panel), $\Sigma_{*}(r)$ (middle panel) and sSFR$(r)$ (right panel). Blue square profile shows an individual radial profile of the sample galaxies, while an average radial profile is shown by green circle profile. The radial profiles are considered up to a semi-major axis of $17$ kpc. Before calculating the average radial profile, each radial profile is extrapolated if it does not reach semi-major axis of $17$ kpc. The extrapolation is done by fitting the radial profile with an exponential function using a least-square fitting method. The fitting is done to the outer region with semi-major axis larger than $3$ kpc to avoid the effect of a bulge component. The extrapolated part of the radial profile is shown with a black line. The error bars in the average radial profiles are calculated using the standard error of mean. 

On average, the $\Sigma_{*}(r)$ and $\Sigma_{\rm SFR}(r)$ of $z\sim 1$ massive disc galaxies have a peak at the centre and gradually decline toward the outskirt, while the average sSFR$(r)$ is almost flat over the entire region. The flat average sSFR$(r)$ agrees with the linear form of the spatially resolved SFMS. We do not see a significant central suppression of sSFR in the average sSFR$(r)$, though it is expected from the flattening trend of the spatially resolved SFMS at high $\Sigma_{*}$.    

\begin{figure*}
\centering
\includegraphics[width=1\textwidth]{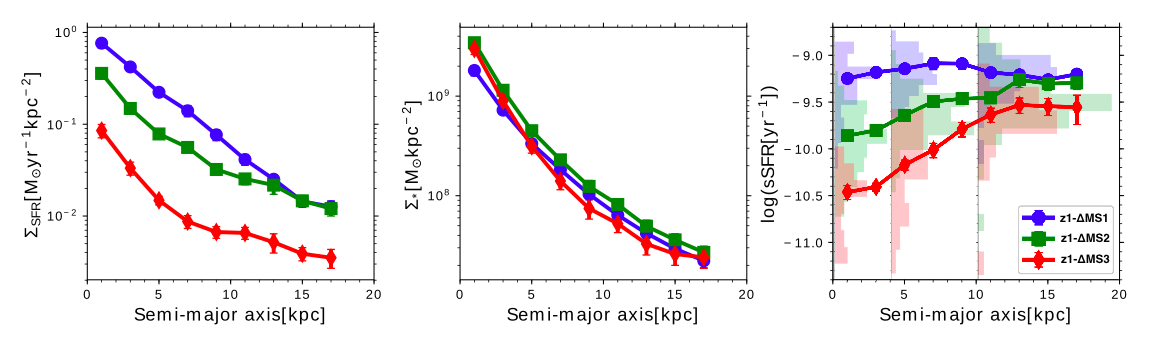}
\caption{Radial profiles of $\Sigma_{\rm SFR}(r)$ (left panel), $\Sigma_{*}(r)$ (middle panel) and sSFR$(r)$ (right panel) as a function of distance from the global SFMS, namely z1-$\Delta$MS1, z1-$\Delta$MS2 and z1-$\Delta$MS3 groups. The radial profiles for the z1-$\Delta$MS1, z1-$\Delta$MS2 and z1-$\Delta$MS3 are shown with blue circles, green squares, and red diamonds with line, respectively. Histograms in the right panel show the sSFR distribution in the central ($r\leqslant 4$ kpc), middle ($4<r\leqslant 10$ kpc) and outskirt ($r>10$ kpc) regions of z1-$\Delta$MS1 (blue), z1-$\Delta$MS2 (green) and z1-$\Delta$MS3 (red) groups.  \label{fig:radial_profile_functionSFMS}}
\end{figure*}

Fig.~\ref{fig:radial_profile_functionSFMS} shows the radial profiles of $\Sigma_{\rm SFR}(r)$ (left panel), $\Sigma_{*}(r)$ (middle panel) and sSFR$(r)$ (right panel) as a function of distance from the global SFMS, namely z1-$\Delta$MS1, z1-$\Delta$MS2 and z1-$\Delta$MS3 groups. On average, the z1-$\Delta$MS1 galaxies have higher $\Sigma_{\rm SFR}$ in all radii than the z1-$\Delta$MS2 galaxies and the z1-$\Delta$MS2 galaxies have higher $\Sigma_{\rm SFR}$ in all radii than the z1-$\Delta$MS3 galaxies. The z1-$\Delta$MS2 and z1-$\Delta$MS3 galaxies have slightly more concentrated $\Sigma_{*}(r)$ with steeper increase toward the central region than the $\Sigma_{*}(r)$ of the z1-$\Delta$MS1 galaxies. The sSFR$(r)$ of those three groups show systematic difference in the central region, while the difference is smaller in the outskirt. The z1-$\Delta$MS1 and z1-$\Delta$MS2 have a sSFR difference of 0.61 dex at semi-major axis of 1 kpc, while the z1-$\Delta$MS1 and z1-$\Delta$MS3 have the sSFR difference of 1.21 dex at the same semi-major axis. The z1-$\Delta$MS1 and z1-$\Delta$MS2 have a sSFR difference of 0.10 dex at semi-major axis of 17 kpc, while the z1-$\Delta$MS1 and z1-$\Delta$MS3 have the sSFR difference of 0.35 dex at the same semi-major axis. Sharp central suppression in the sSFR$(r)$ is observed among the z1-$\Delta$MS2 and z1-$\Delta$MS3 galaxies, while flat sSFR$(r)$ profile is observed for the z1-$\Delta$MS1 galaxies. Those sSFR$(r)$ have correlation with the spatially resolved SFMS of the corresponding groups. The flat sSFR$(r)$ of the z1-$\Delta$MS1 agrees with the linear increasing profile of the spatially resolved SFMS of that group, while the central suppression in the sSFR$(r)$ of the z1-$\Delta$MS2 and z1-$\Delta$MS3 agrees with the flattening trend at high $\Sigma_{*}$ region in the spatially resolved SFMS of those groups. 

To check whether the central suppression in the sSFR$(r)$ profiles of the z1-$\Delta$MS2 and z1-$\Delta$MS3 is real and not caused by a bias toward lower sSFR due to only a few quiescent galaxies, we plot histograms of the sSFR distribution in the central ($r\leqslant 4$ kpc), middle ($4<r\leqslant 10$ kpc) and outskirt ($r>10$ kpc) regions of the z1-$\Delta$MS1 (blue), z1-$\Delta$MS2 (green) and z1-$\Delta$MS3 (red) in the right panel of Fig.~\ref{fig:radial_profile_functionSFMS}. It is shown by the histograms that the sSFRs in the central regions of the z1-$\Delta$MS2 and z1-$\Delta$MS3 are systematically lower than that in the central region of the z1-$\Delta$MS1. It is also shown that the sSFR in all of those three regions of the z1-$\Delta$MS1 have a peak at almost the same sSFR of $\sim 10^{-9.2}yr^{-1}$, which agrees with the flat profile of the sSFR$(r)$ of z1-$\Delta$MS1. Given that dust extinction is increasing toward the central region in massive galaxies \citep[see e.g.][]{nelson2016b,tacchella2018}, one may worry that the centrally suppressed sSFR$(r)$ is actually caused by the red dusty star-forming region which mistakenly recognized as old and passive system. To check if the central regions of the z1-$\Delta$MS2 and z1-$\Delta$MS3 are indeed passive regions, we have calculated the $U$, $V$, and $J$ magnitudes of the galaxies pixels located in the central, middle, and outskirt regions and locate their positions on the $UVJ$ diagrams. We found systematically older and more passive SEDs of pixels located in the central regions of the z1-$\Delta$MS2 and z1-$\Delta$MS3. We discuss this issue in appendix B.

\begin{figure}
\centering
\includegraphics[width=0.5\textwidth]{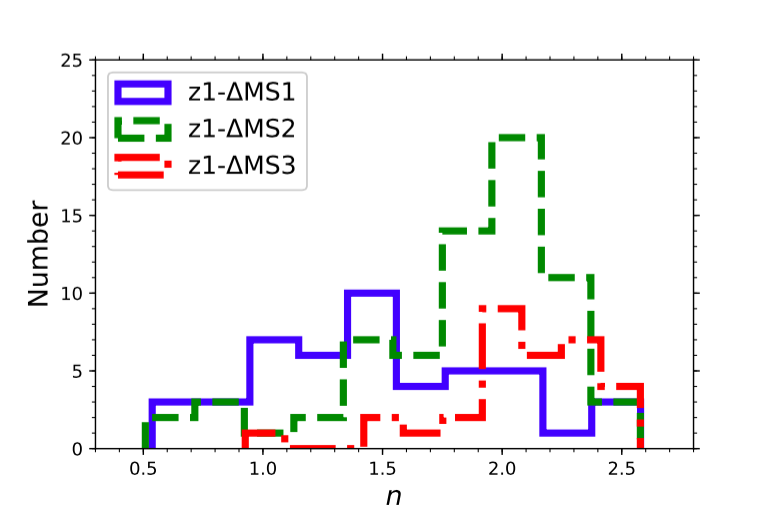}
\includegraphics[width=0.5\textwidth]{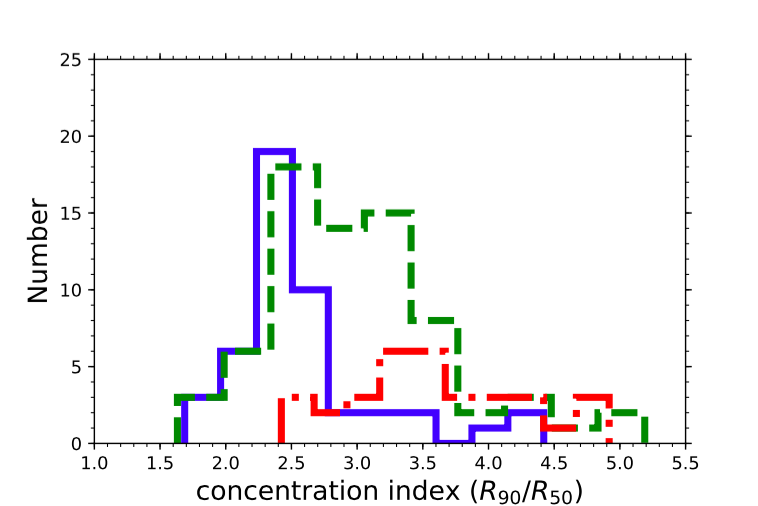}
\caption{Histograms for the distributions of S\' ersic indexes ($n$, top panel) and concentration indexes ($R_{90}/R_{50}$, bottom panel) of z1-$\Delta$MS1 (blue solid line), z1-$\Delta$MS2 (green dashed line), and z1-$\Delta$MS3 (red dashed dotted line). \label{fig:hist_serscidx_concidx}}
\end{figure}

To examine the morphological difference between the z1-$\Delta$MS1, z1-$\Delta$MS2 and z1-$\Delta$MS3, we calculate the S\' ersic index and concentration index ($R_{90}/R_{50}$) of each galaxy in those groups and check the distributions of those properties for the corresponding groups. Fig.~\ref{fig:hist_serscidx_concidx} shows the histograms of the distributions of the S\' ersic indexes ($n$, top panel) and concentration indexes ($R_{90}/R_{50}$, bottom panel). $n$ is calculated by fitting the $\Sigma_{*}(r)$ with S\' ersic profile, $\Sigma_{*}(r)=\Sigma_{*}(r_{0})\exp\left(-\left(\frac{r}{h}\right)^{1/n}\right)$. First, the exponential function ($n=1$) fitting is done to get the initial guess for the radial scale length ($h$) and the zero point ($\Sigma_{*}(r_{0})$). Then the random set of $n$, $h$ and $\Sigma_{*}(r_{0})$ are generated according to the following parameter ranges: $n[0.5,5]$, $\Sigma_{*}(r_{0})[0.1\Sigma_{*}(r_{0},n=1),10\Sigma_{*}(r_{0},n=1)]$ and $h[1,10h_{n=1}]$. The best-fitting S\' ersic profile is determined based on the lowest $\chi^{2}$ value. The $R_{50}$ and $R_{90}$ in the concentration index are calculated with the semi-major axis that enclose 50\% and 90\% of the total $M_{*}$, respectively. In both panels, histogram with blue solid, green dashed and red dashed dotted lines represent the z1-$\Delta$MS1, z1-$\Delta$MS2 and z1-$\Delta$MS3, respectively. The histograms indicate that the z1-$\Delta$MS3 galaxies typically have higher S\' ersic index and concentration index (also higher bulge to total stellar mass ratio, B/T) than the z1-$\Delta$MS1 galaxies, while the z1-$\Delta$MS2 galaxies have both quantities in the intermediate between those two groups. 

Those results suggest an existence of a bulge component in the z1-$\Delta$MS2 and z1-$\Delta$MS3 galaxies, while the z1-$\Delta$MS1 galaxies are disc-dominated. The flat average sSFR$(r)$ profile of the z1-$\Delta$MS1 suggests that those galaxies are still building their stellar mass in the outskirt as well as in the central region. In \citet{abdurrouf2017}, we found that the average sSFR$(r)$ radial profile of the entire $z\sim 0$ sample is centrally suppressed. Those observational results agree with the picture of inside-out quenching where galaxies tend to quench their star formation activities from the central region then the quenching process gradually moves toward the outskirt region. The evidences for the inside-out quenching are also reported by previous research papers, e.g. \citet{tacchella2015}, \citet{gonzalez2016}, \citet{belfiore2018}, \citet{tacchella2018}.    

\section{Discussion}
\label{sec:discussion}

\subsection{Spatially resolved SFMS relations of $z\sim 0$ and $z\sim 1$ samples}
\label{sec:compare_resolved_SFMS}
In order to get insight on the cosmological evolution of the spatially resolved SFMS, we compare the spatially resolved SFMS of the $z\sim 1$ massive disc galaxies with that of the $z\sim 0$ massive disc galaxies. The $z\sim 0$ sample from \citet{abdurrouf2017} is based on $93$ massive face-on disc galaxies at $0.01 < z < 0.02$. Although our selection criteria for the two samples do not guarantee that the $z\sim 0$ sample is the descendant of the $z\sim 1$ sample, it is possible that part of the galaxies from $z\sim 1$ and $z\sim 0$ samples are likely to be on the same evolutionary path, i.e. progenitor and descendant. The comoving volumes covered by $z\sim 1$ and $z\sim 0$ samples are roughly similar ($4.5\times 10^{5} \text{ Mpc}^{3}$ and $4.3\times 10^{5} \text{ Mpc}^{3}$ for the $z\sim 1$ and $z\sim 0$ samples, respectively). However, the median $M_{*}$ of the $z\sim 1$ sample ($7.8\times 10^{10}M_{\odot}$) is systematically higher than that of the $z\sim 0$ sample ($3.5\times 10^{10}M_{\odot}$) and there are $50$ massive disc galaxies ($\log (M_{*}/M_{\odot})\geqslant 11.0$) in the $z\sim 1$ sample, while only $6$ such massive disc galaxies in the $z\sim 0$ sample. A part of the massive disc galaxies at $z\sim 1$ are thought to evolve into elliptical galaxies at $z\sim 0$. The comoving number density ($N$) and stellar mass density ($\rho$) of $\log (M_{*}/M_{\odot})\geqslant 11.0$ disc galaxies in the $z\sim 1$ sample are comparable to those of the elliptical galaxies at $0.01 < z < 0.02$ (taken from MPA-JHU catalog). The comoving number density and stellar mass density of the $z\sim 1$ massive disc galaxies are $\log (N[\text{Mpc}^{-3}])=-3.9$ and $\log (\rho[M_{\odot}\text{Mpc}^{-3}])=7.3$, while those of the local elliptical galaxies are $\log (N[\text{Mpc}^{-3}])=-4.4$ and $\log (\rho[M_{\odot}\text{Mpc}^{-3}])=6.7$, respectively.

\begin{figure}
\centering
\includegraphics[width=0.5\textwidth]{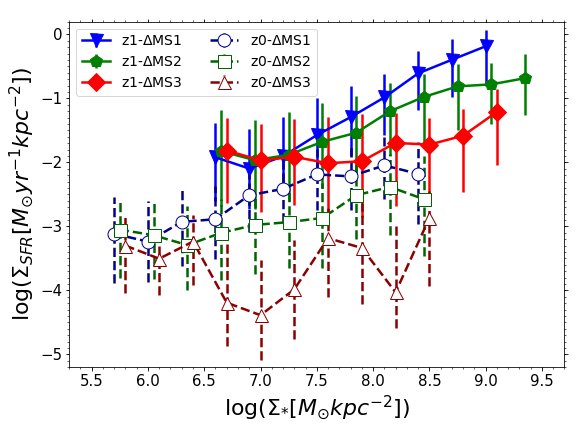}
\caption{Six spatially resolved SFMS relations derived from the $z\sim 1$ sample: z1-$\Delta$MS1 (blue solid tiangles), z1-$\Delta$MS2 (green solid pentagons), and z1-$\Delta$MS3 (red solid diamonds) and $z\sim 0$ sample: z0-$\Delta$MS1 (dark blue open circles), z0-$\Delta$MS2 (dark green open squares), and z0-$\Delta$MS3 (dark red open triangles). The $z\sim 0$ sample from \citet{abdurrouf2017} is based on $93$ massive face-on disc galaxies at $0.01\leqslant z \leqslant 0.02$. For clarity, blue solid triangles and dark blue open circles are shifted by $0.05$ dex to the left, while red solid diamonds and dark red open triangles are shifted by $0.05$ dex to the right from their actual positions. \label{fig:evolution_mode_profile}}
\end{figure}

We compare the spatially resolved SFMS relations of the six groups in the $z\sim 1$ and $z\sim 0$ samples (z1-$\Delta$MS1, z1-$\Delta$MS2, z1-$\Delta$MS3, z0-$\Delta$MS1, z0-$\Delta$MS2 and z0-$\Delta$MS3). The six groups are defined based on the distances from the global SFMS at each redshift. We should emphasize that the classification is based on the order of sSFR at each redshift. In Fig.~\ref{fig:evolution_mode_profile}, the six spatially resolved SFMS relations derived from the $z\sim 0$ and $z\sim 1$ samples are compared. The shift toward higher $\Sigma_{*}$ range for the $z\sim 1$ sample compared to that for the $z\sim 0$ sample is caused by the fact that the $z\sim 1$ sample is systematically more massive than the $z\sim 0$ sample. An obvious feature shown in the Fig.~\ref{fig:evolution_mode_profile} is that the difference in $\Sigma_{\rm SFR}$ at a fixed $\Sigma_{*}$ between the two spatially resolved SFMS relations at low $\Sigma_{*}$ region is smaller than that at high $\Sigma_{*}$ region. If we quantitatively compare the spatially resolved SFMS of galaxies in the highest sSFRs groups, i.e. z1-$\Delta$MS1 and z0-$\Delta$MS1, the sSFR difference is $0.4$ dex at $\log(\Sigma_{*}[M_{\odot} \text{kpc}^{-2}])=7.0$, while that is $1.5$ dex at $\log(\Sigma_{*}[M_{\odot} \text{kpc}^{-2}])=8.5$. This trend suggests that the star formation activity in the disc region (represented with low $\Sigma_{*}$ value) shows less suppression from $z\sim 1$ to $z\sim 0$ compared to the star formation activity in the central region (represented with high $\Sigma_{*}$ value). The trend agrees with the inside-out quenching scenario \citep[e.g.][]{tacchella2015,gonzalez2016,belfiore2018,tacchella2018}.

\subsection{Empirical model for the evolution of \texorpdfstring{$\Sigma_{*}(r)$}{Lg}, \texorpdfstring{$\Sigma_{\rm SFR}(r)$}{Lg} and sSFR\texorpdfstring{$(r)$}{Lg} radial profiles at \texorpdfstring{$0\lesssim z \lesssim1$}{Lg}}
\label{sec:empirical_model}

We try to construct an empirical model for the evolution of the radial profiles of $\Sigma_{\rm SFR}(r)$, $\Sigma_{*}(r)$ and sSFR$(r)$ at $0\lesssim z\lesssim 1$, by defining possible pairs of the progenitor and descendant galaxies from the $z\sim 1$ and $z\sim 0$ samples, based on the location on the global SFMS. We define the pairs as follows: (1) we start from a galaxy at $z=2$ that has sSFR and $M_{*}$ within $\pm 0.3$ dex from the global SFMS relation of \citet{speagle2014} at that epoch, and use the sSFR and $M_{*}$ as a starting point for drawing a galaxy evolutionary track in the $\log (M_{*})$-$\log (\text{sSFR})$ plane. (2) The star formation history (SFH) of the galaxy is assumed to be in the exponentially declining form, $\text{SFR}(t)=\text{SFR}(t_{0})e^{-\Delta t/\tau}$ with $t=t_{0}+\Delta t$ and $t_{0}$ as the age of the universe at $z=2$. (3) We choose a set of model parameters (which are $M_{*}(t_{0})$, sSFR$(t_{0})$ and $\tau$) which can select as many galaxies as possible from the $z\sim 1$ and $z\sim 0$ samples so that the model evolutionary track can be a possible evolutionary path connecting the two samples. 

Using the average $\Sigma_{\rm SFR}(r)$ of the progenitor and descendant samples selected from the above assumptions, we can infer the radially-resolved SFH, from which we can follow the radial stellar mass buildup during the epoch of $0\lesssim z \lesssim 1$. We consider three different evolutionary paths: two with long and short $\tau$, and the other one with just consider a same mass range. To make a model evolutionary track, we assume a certain range for each parameter which produces broad evolutionary track, instead of assuming a single value for each model parameter which only produces an evolutionary track with a single line. The two models with exponentially declining SFH are: (a) model with parameter ranges of $\log (M_{*}(t_{0}))=[9.7:9.9]$, $\log (\text{sSFR}(t_{0}))=[-8.6:-8.4]$ and $\tau = [4.0:6.0]$, hereafter called model A; and (b) model with $\log (M_{*}(t_{0}))=[10.2:10.3]$, $\log (\text{sSFR}(t_{0}))=[-8.7:-8.5]$ and $\tau = [1.3:2.5]$, hereafter called model B. The $M_{*}$, sSFR and $\tau$ are in unit of $M_{\odot}$, $\text{yr}^{-1}$ and Gyr, respectively. The third model, which is called model C, is made without any assumption on the SFH and only connects galaxies in the stellar mass range of $10.85\leq \log (M_{*}/M_{\odot}) \leq 11.2$.

\begin{figure*}
\centering
\includegraphics[width=0.32\textwidth]{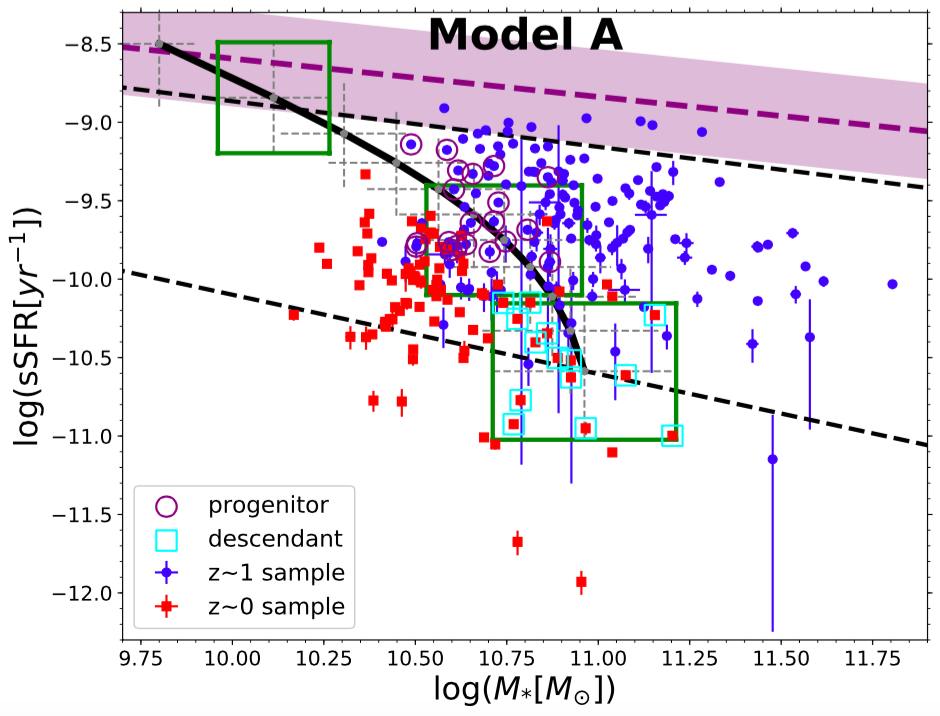}
\includegraphics[width=0.32\textwidth]{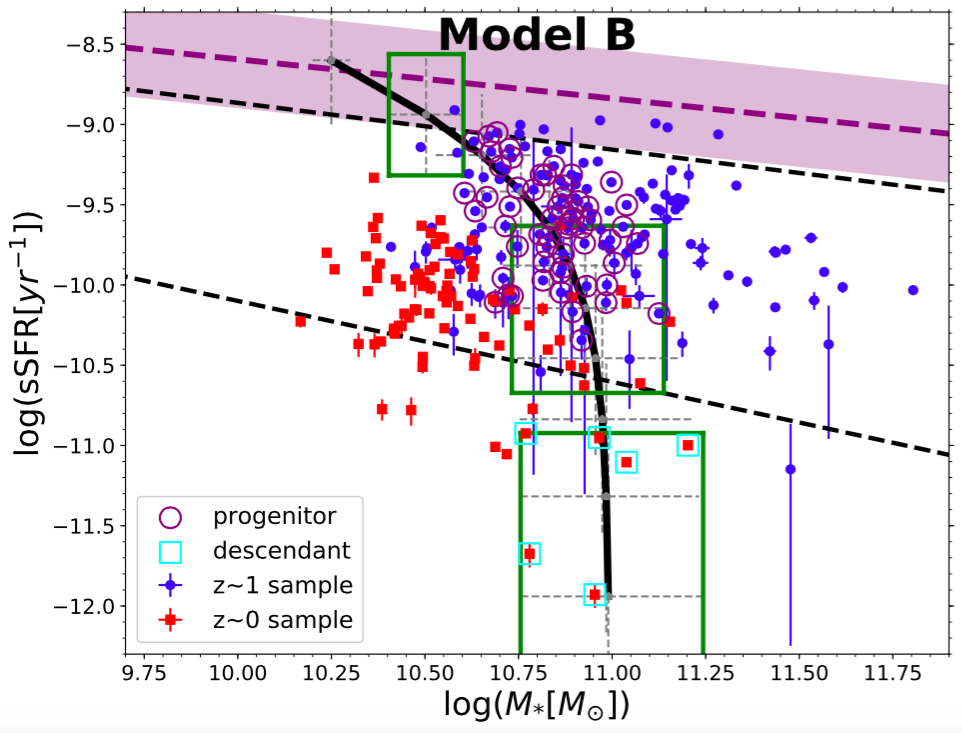}
\includegraphics[width=0.32\textwidth]{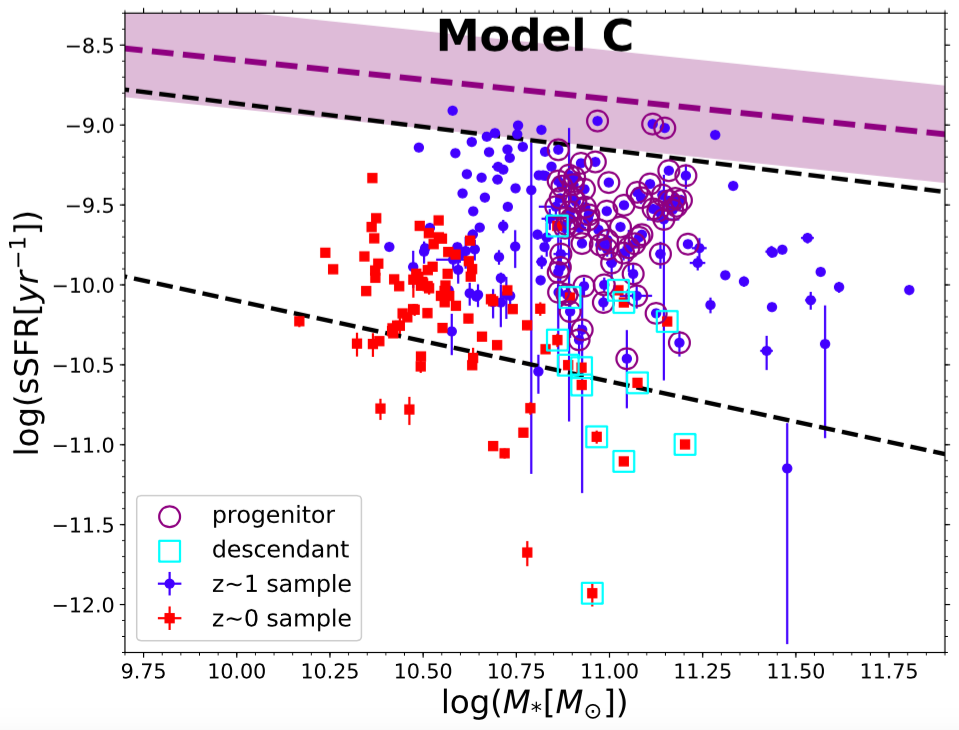}
\caption{Left panel: the evolutionary track and the selected progenitors and descendants of model A which has $\log (M_{*}(t_{0}))=[9.7:9.9]$, $\log (\text{sSFR}(t_{0}))=[-8.6:-8.4]$ and $\tau = [4.0:6.0]$. Middle panel: the evolutionary track and the selected progenitors and descendants of model B which has $\log (M_{*}(t_{0}))=[10.2:10.3]$, $\log (\text{sSFR}(t_{0}))=[-8.7:-8.5]$ and $\tau = [1.3:2.5]$. The $M_{*}$, sSFR and $\tau$ are in unit of $M_{\odot}$, $\text{yr}^{-1}$ and Gyr, respectively. The black lines in both panels represent the model evolutionary tracks if the middle value of each model parameter range is used. The gray dashed-line show the range of sSFR and $M_{*}$ if the model parameter ranges are considered. Three green boxes show the ranges of sSFR and $M_{*}$ given by the horizontal and vertical "errorbars" of the model evolutionary track calculated at $z=1.8$, $0.8$, and $0$. Right panel: the selected progenitors and descendants of model C that is made without any assumption on the SFH, except the mass range of $10.85\leqslant \log (M_{*}/M_{\odot}) \leqslant 11.2$. The purple dashed line and purple shaded area represent the global SFMS relation at $z=2$ and $\pm 0.3$ dex scatter around it, respectively. The black dashed-lines represent the global SFMS relations at $z=1.2$ and $z=0.015$.
\label{fig:estimate_evolution_galaxies_localdesc_highzprog}}
\end{figure*}

Fig.~\ref{fig:estimate_evolution_galaxies_localdesc_highzprog} shows the model evolutionary tracks and the selected progenitor and descendant galaxies for the model A (left panel), B (middle panel) and C (right panel). The black lines represent the model evolutionary tracks if the model parameters are taken from the middle values of the ranges, while the vertical and horizontal gray dashed-lines at each redshift represent the ranges of the sSFR and $M_{*}$ if the model parameter ranges are used. The vertical 'error bar' is extended by $0.3$ dex above and below from the actual length to make it roughly as wide as the scatter of the global SFMS (which is expected to be able to account for a fluctuations of a real galaxy evolutionary path around the simple exponentially decaying form), while the horizontal 'error bar' is kept as the original length. The scatter in the vertical direction also accounts for the higher uncertainty of the sSFR compared to $M_{*}$ of the sample galaxies. The progenitors (descendants) are defined as the galaxies from $z\sim 1$ ($z\sim 0$) sample which are enclosed within the 'box' given by the vertical and horizontal 'errorbars', evaluated at the redshifts of the galaxies. Three green boxes show the ranges in sSFR and $M_{*}$ given by the horizontal and vertical 'error bars' of the model evolutionary track calculated at $z=1.8$, $0.8$, and $0$. The number of progenitors (descendants) selected using the model A, B and C are $20 (14)$, $57 (6)$ and $71 (14)$, respectively. As expected from the larger value of $\tau$, the sSFR of model A decline more slowly compared to that of model B. The purple dashed line and purple shaded region represent the global SFMS relation at $z=2$ and $\pm 0.3$ dex scatter around it, respectively. The black dashed-lines represent the global SFMS relations at $z=1.2$ and $z=0.015$.     

\begin{figure*}
\centering
\includegraphics[width=0.95\textwidth]{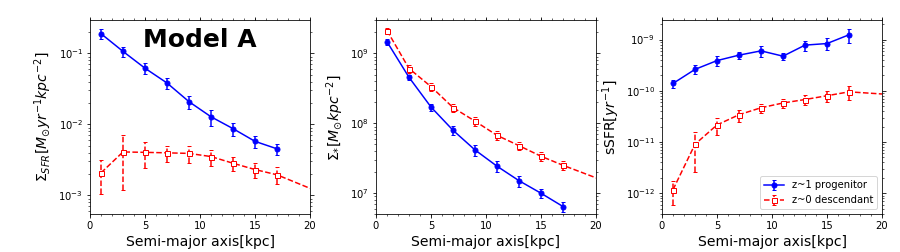}
\includegraphics[width=0.95\textwidth]{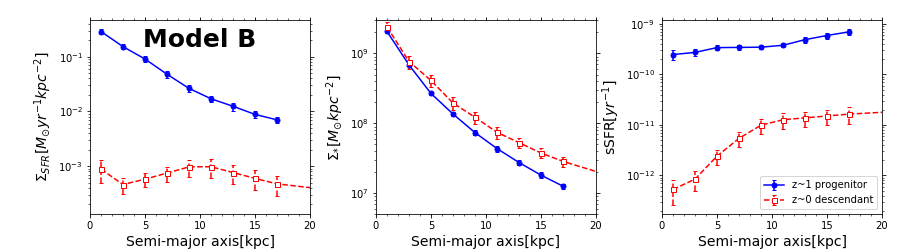}
\includegraphics[width=0.95\textwidth]{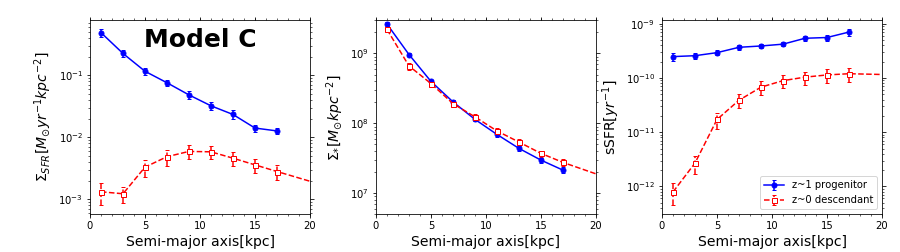}
\caption{Average $\Sigma_{\rm SFR}(r)$ (left panel in each row), $\Sigma_{*}(r)$ (middle panel in each row) and sSFR$(r)$ (right panel in each row) radial profiles of the progenitors (blue circles with solid line) and descendants (red open squares with dashed line) selected by model A (first row), model B (second row) and model C (third row).  \label{fig:radial_profiles_prog_desc_combined}}
\end{figure*}

Fig.~\ref{fig:radial_profiles_prog_desc_combined} shows the average radial profiles of the selected progenitors (blue circle with solid line) and descendants (red open square with dashed line) galaxies using the evolutionary tracks of the model A (first row), B (second row) and C (third row). The average radial profiles of $\Sigma_{\rm SFR}(r)$ and sSFR$(r)$ show that the star formation activity is declined in all radii from $z\sim 1$ to $z\sim 0$ with larger decline in the central region compared to that in the outskirt. The stellar mass buildup in model A shows larger stellar mass increase over all radii compared to that in model B, as expected from the larger $\tau$ of model A than that of model B. The radial stellar mass increase is not found in model C. 

Given the radial decrease of $\Sigma_{\rm SFR}(r)$ from $z\sim 1$ to $z\sim 0$, we derive an empirical model for the evolution of the $\Sigma_{\rm SFR}(r)$, $\Sigma_{*}(r)$ and sSFR$(r)$. Here, we assume exponentially declining SFH at each radius in the form
\begin{equation}
\Sigma_{\rm SFR}(r,t) = \Sigma_{\rm SFR}(r,t_{0}) e^{-\Delta t/\tau(r)}
\label{eq:radial_SFH}
\end{equation}
where $t=t_{0}+\Delta t$ with $t_{0}$ is the age of the universe at the median redshift of the progenitors. The median redshift of the selected progenitors (descendants) by model A, B and C are $1.064\pm 0.026$ ($0.016\pm 0.001$), $1.133\pm 0.043$ ($0.017\pm 0.002$) and $1.216\pm 0.044$ ($0.017\pm 0.002$), respectively. The uncertainty of the median redshift (which is calculated using bootstrap resampling method) is used in later analysis for calculating the uncertainty of model properties, such as radial profile of SFH, $\Sigma_{*}(r)$ and sSFR$(r)$.  

\begin{figure*}
\centering
\includegraphics[width=0.95\textwidth]{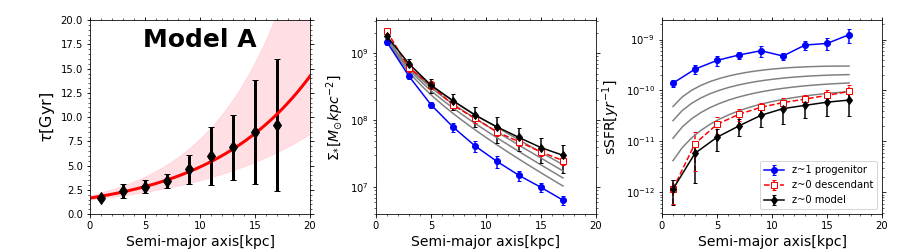}
\includegraphics[width=0.95\textwidth]{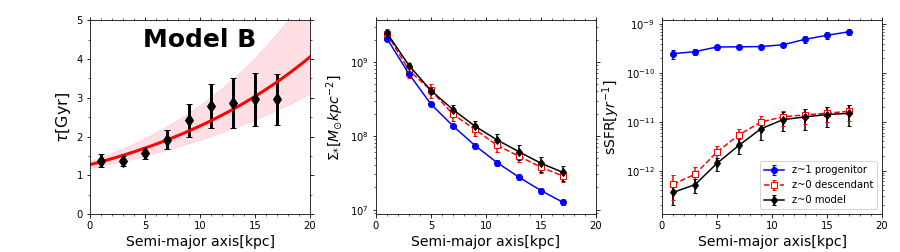}
\includegraphics[width=0.95\textwidth]{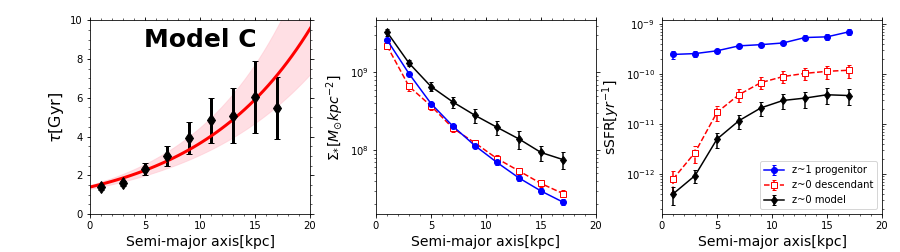}
\caption{Comparison between average $\Sigma_{*}(r)$ and sSFR$(r)$ radial profiles of the descendant galaxies and those radial profiles predicted by the empirical models at $z\sim 0$. The first, second, and third rows show the radial profile of $\tau(r)$ (left panel in each row), the observed and predicted radial profiles of $\Sigma_{*}(r)$ (middle panel in each row) and the observed and predicted radial profiles of sSFR$(r)$ (right panel in eah row) for model A, B and C, respectively. The black diamonds in the left panel in each row show the $\tau(r)$ of each model, while the red line and red-shaded region around it show the best-fitting exponential function of the $\tau(r)$ and its $1\sigma$ uncertainty, respectively. The $\Sigma_{*}(r)$ and sSFR$(r)$ for progenitor, descendant, and model prediction are shown with blue closed circles with solid line, red open squares with dashed line, and black diamonds with solid line, respectively. \label{fig:radial_profile_predicted_empiricalmodel} }
\end{figure*}

Using Eq.~\ref{eq:radial_SFH} with $\Delta t$ as the time difference between the median redshifts of the progenitors and descendants ($7.74\pm 0.10$, $7.97\pm 0.17$ and $8.25\pm 0.12$ Gyr for model A, B and C, respectively), we calculate the $\tau$ at each radius. The results for all three models are shown in the left panel in each row of Fig.~\ref{fig:radial_profile_predicted_empiricalmodel}. The $\tau(r)$ is increasing with increasing radius in all three models. The errorbar at each radius is the $1 \sigma$ uncertainty calculated through a Monte-Carlo method, which calculate $\tau(r)$ randomly by varying the average $\Sigma_{\rm SFR}(r)$ of the progenitors and descendants and $\Delta t$ within their uncertainties following Gaussian distribution. The uncertainty of $\Delta t$ is calculated using a Monte-Carlo method, which calculate $\Delta t$ randomly by varying the median redshifts of the progenitors and descendants within their uncertainties following Gaussian distribution. The red line in the $\tau(r)$ plot shows the result of exponential function fitting and the red shaded area shows its $1\sigma$ uncertainty. They are calculated using a Bayesian statistic method. The middle and right panels in each row show the predicted $\Sigma_{*}(r)$ and sSFR$(r)$ by the model at the median redshift of the descendants (shown with a black line). The predicted $\Sigma_{*}(r)$ and sSFR$(r)$ by the model A and B are consistent with the average radial profiles of the descendants, while those of model C show large discrepancy from the observed radial profiles at $z\sim 0$. The consistency suggests that model A and B are possible evolutionary models describing the radial stellar mass accumulation in massive disc galaxies. The simple exponentially declining radial SFH model can explain the stellar mass buildup by the star formation activity in the massive disc galaxies.        

Mathematical descriptions of the evolution of the $\Sigma_{\rm SFR}(r)$, $\Sigma_{*}(r)$ and sSFR$(r)$ radial profiles are constructed based on the model A. At first, the average $\Sigma_{\rm SFR}(r)$ and $\Sigma_{*}(r)$ radial profiles of the progenitors are fitted with exponential function and S\' ersic profile, respectively, and the best-fitting profiles are used as the initial condition from which the radial profiles at subsequent times are calculated. Those fitting results are  

\begin{equation}
\Sigma_{\rm SFR}(r,t_{0}) = (0.21 \pm 0.03)e^{-r/(4.18 \pm 0.24)},
\label{eq:fit_SFR_profile}
\end{equation}

\begin{equation}
\Sigma_{*}(r,t_{0}) = (8.43\times 10^{9} \pm 4.43\times 10^{8})e^{-\left(\frac{r}{0.35 \pm 0.02}\right)^{\left(\frac{1}{1.96 \pm 0.03}\right)}},
\label{eq:fit_SM_profile}
\end{equation}
The time scale of star formation at each radius is determined by an exponential function fitting to the $\tau(r)$ as

\begin{equation}
\tau(r) = (1.66 \pm 0.22)e^{r/\left(9.32 \pm 2.21\right)},
\label{eq:fit_tau_profile}
\end{equation}
The best-fitting exponential function is shown in the left panel of the first row of Fig.~\ref{fig:radial_profile_predicted_empiricalmodel} with a red line. The mathematical prescription for the radial profile evolutions are as follows

\begin{equation}
\Sigma_{\rm SFR}(r,t) = \Sigma_{\rm SFR}(r,t_{0}) e^{-(t-t_{0})/\tau(r)},
\end{equation}

\begin{equation}
\Sigma_{*}(r,t) = \Sigma_{*}(r,t_{0}) + \tau(r) \Sigma_{\rm SFR}(r,t_{0}) \left(1 - e^{-(t-t_{0})/\tau(r)}\right),
\label{eq:SM_empirical_model}
\end{equation}
where $t_{0}$ is the age of the universe at the median redshift of the progenitors and $t$ is the cosmic time within $0\lesssim z \lesssim 1$. The $\Sigma_{*}(r)$ and sSFR$(r)$ at $z=0.8$, $0.6$, $0.4$ and $0.2$ calculated based on the above empirical model are shown as gray lines in the middle and right panels of the first row in Fig.~\ref{fig:radial_profile_predicted_empiricalmodel}. The empirical model for the evolution of the $\Sigma_{*}(r)$ shows stellar mass buildup in inside-to-outside manner. This inside-out stellar mass buildup in the galaxies is also found by previous researches e.g. \citet{vandokkum2010, nelson2016, morishita2015, tacchella2015, tadaki2017}.  

\begin{figure}
\centering
\includegraphics[width=0.5\textwidth]{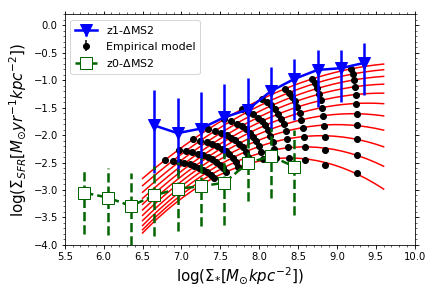}
\caption{Evolution of the spatially resolved SFMS relation as a function of redshift inferred by the empirical model of $\Sigma_{\rm SFR}(r)$ and $\Sigma_{*}(r)$ radial profiles. The spatially resolved SFMS relations at $0.11$ redshift steps between $z=0$ and $z=1.1$ constructed using the empirical model are shown with black circles. The red line on each spatially resolved SFMS represents best-fitting second order polynomial function. The blue triangles with solid line and green squares with dashed line represent observed spatially resolved SFMS of z1-$\Delta$MS2 and z0-$\Delta$MS2, respectively.  \label{fig:SFMS_from_radial_profile}}
\end{figure}

We check the consistency between the empirical model of $\Sigma_{\rm SFR}(r,t)$ and $\Sigma_{*}(r,t)$ radial profiles and the spatially resolved SFMS at $z\sim 0$ and $z\sim 1$. Fig.~\ref{fig:SFMS_from_radial_profile} shows the spatially resolved SFMS relations at redshift interval of $0.11$ between $0\leqslant z \leqslant 1.1$ (black circles) constructed from the empirical model of radial profiles. The red lines represent the best-fitting second order polynomial functions to the spatially resolved SFMS constructed from the empirical model. The blue triangles and green squares represent the observed spatially resolved SFMS relations of the z1-$\Delta$MS2 and z0-$\Delta$MS2 galaxies, respectively. The observed spatially resolved SFMS from those two groups are used for the comparison because large fraction of the progenitor and descendant galaxies are belong to those groups. The spatially resolved SFMS relations at $z=1.1$ and $z=0$ predicted by the empirical model agree with the observed spatially resolved SFMS of z1-$\Delta$MS2 and z0-$\Delta$MS2, respectively.

\subsection{The radial quenching timescale derived from the empirical model}

\begin{figure}
\centering
\includegraphics[width=0.5\textwidth]{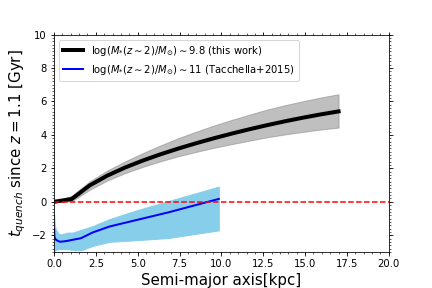}
\caption{Radial profile of the quenching timescale ($t_{\rm quench}(r)$) from the age of the universe at $z=1.1$. Negative time corresponds to the cosmic time at $z>1.1$. The black line represents $t_{\rm quench}(r)$ obtained from this work for $10.5<\log (M_{*}/M_{\odot})<10.9$ galaxies at $z=1.1$, which corresponds to the low-mass galaxies ($9.7\lesssim \log (M_{*}/M_{\odot})\lesssim 9.9$) at $z=2$ (according to model A, see Section~\ref{sec:empirical_model}). The blue line represents $t_{\rm quench}(r)$ profile reported by \citet{tacchella2015} for massive galaxies, with stellar mass range of $10.8\leqslant \log (M_{*}/M_{\odot})<11.7$ at $z\sim 2$. \label{fig:radial_profile_quenching_time_longtau}}
\end{figure}

In this section, we estimate the quenching timescale at each radius to quantitatively examine the inside-out quenching process of the sample galaxies. Using the empirical model derived in the previous section, we derive the radial profile of the quenching timescale ($t_{\rm quench}(r)$). The quenching timescale is assumed to be the time needed for the sSFR in each radius ($\Sigma_{\rm SFR}(r,t)/\Sigma_{*}(r,t)$) to reach a critical value of $10^{-10}yr^{-1}$, which is also used to separate star-forming and quiescent galaxies by \citet{peng2010} and star-forming and quiescent sub-galactic region by \citet{gonzalez2016}, which corresponds to the mass doubling time of $10$ Gyr, i.e. larger than the Hubble time at $z\gtrsim 0.5$. Black line in Fig.~\ref{fig:radial_profile_quenching_time_longtau} shows $t_{\rm quench}(r)$ from $z=1.1$. The gray shaded area around the line represents the $1\sigma$ uncertainty calculated using the Monte-Carlo method which is done by randomly varying all the parameters involved in the calculation ($\Sigma_{\rm SFR}(r,t0)$, $\Sigma_{*}(r,t0)$ and $\tau(r)$) within their uncertainties by assuming Gaussian distribution, then calculate the standard deviation of the $t_{\rm quench}$ at each radius.  

Inside-out quenching process is clearly shown by the $t_{\rm quench}(r)$ profile. The $t_{\rm quench}(r)$ shows that the central regions ($r\sim 1$ kpc) will quench by $\sim 200$ Myr from $z=1.1$, while the outskirt ($r \sim 15$ kpc) will quench by $\sim 5.2$ Gyr from $z=1.1$. The model A from which the empirical model is derived has initial mass at $z=2$ of $9.7\leqslant \log (M_{*}/M_{\odot}) \leqslant 9.9$ and the progenitor galaxies selected using this model have $10.5<\log (M_{*}/M_{\odot})<10.9$ at $z\sim 1.1$. The blue profile in Fig.~\ref{fig:radial_profile_quenching_time_longtau} represents the $t_{\rm quench}(r)$ reported by \citet{tacchella2015} for very massive galaxies with stellar mass range of $10.8\leqslant \log (M_{*}/M_{\odot})<11.7$, at $z\sim 2$, which has been subtracted by the cosmic time interval between $z=1.1$ and $z=2.2$. The $t_{\rm quench}(r)$ profile of \citet{tacchella2015} is derived based on the average $\Sigma_{*}(r)$ and $\Sigma_{\rm SFR}(r)$ of massive galaxies at $z\sim 2$ and the average $\Sigma_{*}(r)$ of similarly massive early-type galaxies at $z\sim 0$. By assuming that the $z\sim 2$ galaxies keep forming stars with their observed $\Sigma_{\rm SFR}(r)$, they estimated the time needed for each radius to stop their star formation in order not to overshoot the $\Sigma_{*}(r)$ of the $z\sim 0$ galaxies. By the calculation, they shown that the integrated SFR at any given time is following that of typical main-sequence galaxies.

The blue $t_{\rm quench}(r)$ shows the inside-out quenching process of the $z\sim 2$ massive galaxies, of which the central region is quenched since $z\sim 2$, and their star formation is fully quenched in the entire region by $z\sim 1$. The $t_{\rm quench}(r)$ of low mass (this work) and very massive galaxies \citep{tacchella2015} are differ in a starting time of the quenching in the central region, while their slopes are similar. Those $t_{\rm quench}(r)$ trends agree with the "downsizing" scenario \citep[e.g.][]{cowie1996, Juneau2005} and furthermore suggests that the "downsizing" phenomenon appear even in the spatially resolved properties. The massive galaxies tend to quench faster in all radii than the low mass galaxies. \citet{perez2013} also found the indication that the "downsizing" phenomenon is spatially preserved by analyzing the spatially resolved stellar mass assembly history in local galaxies using integral field spectroscopy observation. They found that massive galaxies assemble their stellar mass faster than low mass galaxies in both inner and outer regions.  

\section{Summary}

We investigate the relation between local surface density (at the $\sim 1$kpc scale) of SFR ($\Sigma_{\rm SFR}$) and stellar mass ($\Sigma_{*}$), so-called spatially resolved SFMS, in the massive ($\log(M_{*}/M_{\odot})>10.5$) face-on disc galaxies at $0.8 < z < 1.8$ and located in the GOODS-S region. We also study the radial profiles of $\Sigma_{\rm SFR}(r)$, $\Sigma_{*}(r)$ and sSFR$(r)$. The effect of the integrated sSFR to the spatially resolved SFMS and the radial profiles of $\Sigma_{\rm SFR}(r)$, $\Sigma_{*}(r)$ and sSFR$(r)$ are discussed. By employing our previous results for $z\sim 0$ massive ($\log(M_{*}/M_{\odot})>10.5$) face-on disc galaxies \citep{abdurrouf2017}, we discuss the evolution of the spatially resolved SFMS and the radial profiles of $\Sigma_{\rm SFR}(r)$, $\Sigma_{*}(r)$ and sSFR$(r)$ during the epoch of $0\lesssim z \lesssim 1$. 

To derive the spatially resolved SFR and stellar mass of a galaxy at $z\sim 1$, we use a method so-called pixel-to-pixel SED fitting, which fits the spatially resolved photometric SED in each bin of a galaxy to the library of model photometric SEDs using the Bayesian statistics approach. The spatially resolved SED of a galaxy with rest-frame FUV-NIR coverage is constructed using 8 bands imaging data from CANDELS and 3D-HST. 

Our results can be summarized as follows.

\begin{enumerate}[leftmargin=*]
	\item[1.] We find the relation between $\Sigma_{\rm SFR}$ and $\Sigma_{*}$, so-called spatially resolved SFMS, in the $z\sim 1$ sample. This relation has a linear form with the slope of $1.01$ in the galaxies which lie within $\pm 0.3$ dex from the global SFMS (i.e. z1-$\Delta$MS1), while a flattening trend at high $\Sigma_{*}$ end is observed in the spatially resolved SFMS of galaxies which lie between $-0.3$ and $-0.8$ dex (i.e. z1-$\Delta$MS2) and below $-0.8$ dex (i.e. z1-$\Delta$MS3) from the global SFMS. 

	\item[2.] The sSFR$(r)$ radial profiles of the z1-$\Delta$MS2 and z1-$\Delta$MS3 galaxies show decline in the central region, while sSFR$(r)$ radial profile of the z1-$\Delta$MS1 is flat over the entire radius. The central suppression in the sSFR$(r)$ radial profiles of the z1-$\Delta$MS2 and z1-$\Delta$MS3 corresponds to the flattening at high $\Sigma_{*}$ end of the spatially resolved SFMS of the corresponding groups. Morphology of the z1-$\Delta$MS3 galaxies show higher S\' ersic index and concentration index ($R_{90}/R_{50}$) compared to those of the z1-$\Delta$MS1, while the S\' ersic index and concentration index of the z1-$\Delta$MS2 galaxies are in the intermediate between those two groups. This trend suggests the existence of central bulge components in the z1-$\Delta$MS2 and z1-$\Delta$MS3 galaxies, while z1-$\Delta$MS1 galaxies are disc-dominated system and still building their stellar mass in both of the central region and outskirt.

	\item[3.] The spatially resolved SFMS shows smaller decline (i.e. smaller decrease of sSFR=$\Sigma_{\rm SFR}/\Sigma_{*}$) in the low $\Sigma_{*}$ region than that in the high $\Sigma_{*}$ region from $z\sim 1$ to $z\sim 0$. This trend suggests that the star formation rate in the disc region experienced less suppression compared to the star formation rate in the central region during that epoch, agrees with the inside-out quenching scenario.
	
	\item[4.] By selecting pairs of possible progenitors and descendants from the $z\sim 1$ and $z\sim 0$ samples using model evolutionary track with exponentially declining SFH, and then using the average $\Sigma_{\rm SFR}(r)$ of the progenitor and descendant galaxies to obtain the radially-resolved SFH following exponentially declining form, we derive the empirical model for the evolution of the $\Sigma_{\rm SFR}(r)$, $\Sigma_{*}(r)$ and sSFR$(r)$ radial profiles. The empirical model successfully reproduces the observed $\Sigma_{*}(r)$ and sSFR$(r)$ radial profiles at $z\sim 0$ and also consistent with the spatially resolved SFMS at $z\sim 1$ and $z\sim 0$. 
	
	\item[5.] Using the empirical model for the evolution of the $\Sigma_{\rm SFR}(r)$ and $\Sigma_{*}(r)$, we estimate the radial profile of the quenching timescale. $t_{\rm quench}(r)$ is increasing with increasing radius which shows an inside-out progression of the quenching process of the sample galaxies. The quenching timescale at each radius is later than that reported by \citet{tacchella2015} for more massive galaxies. This result suggests that "downsizing" signal is spatially preserved i.e. faster quenching of massive galaxies than low mass galaxies in the entire radius.

\end{enumerate}   

\section*{Acknowledgements}
We thanks anonymous referee for his/her comments which improve our paper. We thanks Drs. Takahiro Morishita and Sandro Tacchella for their useful comments. We thanks Dr. Sandro Tacchella for providing the radial profile of quenching timescale of massive galaxies at $z\sim 2$. Abdurro'uf acknowledges the support from Japanese Government (MEXT) scholarship for his studies. 
   
This work is based on observations taken by the 3D-HST Treasury Program (GO 12177 and 12328) with the NASA/ESA HST, which is operated by the Association of Universities for Research in Astronomy, Inc., under NASA contract NAS5-26555. This work is based on observations taken by the CANDELS Multi-Cycle Treasury Program with the NASA/ESA HST, which is operated by the Association of Universities for Research in Astronomy, Inc., under NASA contract NAS5-26555.

This work is based on observations made with the NASA \textit{Galaxy Evolution Explorer}. GALEX is operated for NASA by the California Institute of Technology under NASA contract NAS5-98034. This work has made use of SDSS data. Funding for the Sloan Digital Sky Survey IV has been provided by the Alfred P. Sloan Foundation, the U.S. Department of Energy Office of Science, and the Participating Institutions. SDSS-IV acknowledges support and resources from the Center for High-Performance Computing at the University of Utah. The SDSS web site is www.sdss.org. SDSS is managed by the Astrophysical Research Consortium for the 
Participating Institutions of the SDSS Collaboration including the 
Brazilian Participation Group, the Carnegie Institution for Science, 
Carnegie Mellon University, the Chilean Participation Group, the French Participation Group, Harvard-Smithsonian Center for Astrophysics, 
Instituto de Astrof\'isica de Canarias, The Johns Hopkins University, 
Kavli Institute for the Physics and Mathematics of the Universe (IPMU) / 
University of Tokyo, Lawrence Berkeley National Laboratory, 
Leibniz Institut f\"ur Astrophysik Potsdam (AIP),  
Max-Planck-Institut f\"ur Astronomie (MPIA Heidelberg), 
Max-Planck-Institut f\"ur Astrophysik (MPA Garching), 
Max-Planck-Institut f\"ur Extraterrestrische Physik (MPE), 
National Astronomical Observatories of China, New Mexico State University, 
New York University, University of Notre Dame, 
Observat\'ario Nacional / MCTI, The Ohio State University, 
Pennsylvania State University, Shanghai Astronomical Observatory, 
United Kingdom Participation Group,
Universidad Nacional Aut\'onoma de M\'exico, University of Arizona, 
University of Colorado Boulder, University of Oxford, University of Portsmouth, 
University of Utah, University of Virginia, University of Washington, University of Wisconsin, 
Vanderbilt University, and Yale University.

%%%%%%%%%%%%%%%%%%%%%%%%%%%%%%%%%%%%%%%%%%%%%%%%%%



\bibliographystyle{mnras}
\bibliography{mybib} % if your bibtex file is called example.bib

%\begin{thebibliography}{99}
%\end{thebibliography}

%%%%%%%%%%%%%%%%%%%%%%%%%%%%%%%%%%%%%%%%%%%%%%%%%%

%%%%%%%%%%%%%%%%% APPENDICES %%%%%%%%%%%%%%%%%%%%%

\appendix

\section{Comparison between $M_{*}$ and SFR derived from spatially-resolved and global SED fitting}
\begin{figure}
\centering
\includegraphics[width=0.45\textwidth]{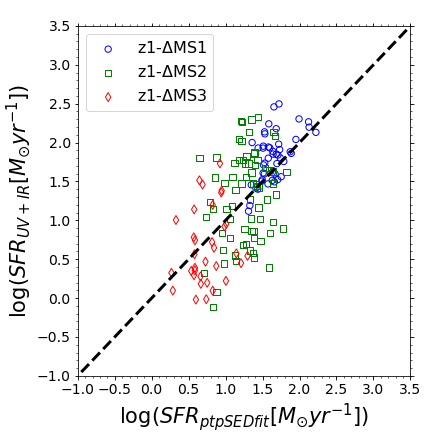}
\includegraphics[width=0.45\textwidth]{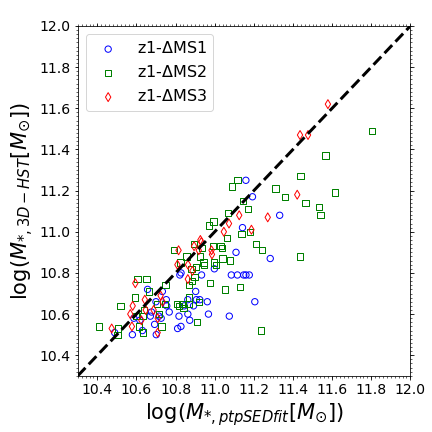}
\caption{Top panel: same as Fig.~\ref{fig:compare_SFR_ptpSEDfit_3DHST} but now distribution of the sub-samples based on the distance from the global SFMS is shown in the comparison plot. Bottom panel: comparison between $M_{*}$ estimated from the pixel-to-pixel SED fitting method ($M_{*,\text{ptpSEDfit}}$) and the $M_{*}$ taken from the 3D-HST catalog ($M_{*,\text{3D-HST}}$). Blue circles, green squares, and red diamonds represent z1-$\Delta$MS1, z1-$\Delta$MS2, and z1-$\Delta$MS3 sub-samples, respectively. \label{SFR_SM_3DHST_vs_ptpSEDfit} }
\end{figure}

Top panel of Fig.~\ref{SFR_SM_3DHST_vs_ptpSEDfit} shows distributions of the z1-$\Delta$MS1, z1-$\Delta$MS2, and z1-$\Delta$MS3 galaxies on a comparison plot between the SFR estimated using our method ($\text{SFR}_\text{ptpSEDfit}$) and the SFR taken from the 3D-HST catalog ($\text{SFR}_\text{UV+IR}$), same as Fig.~\ref{fig:compare_SFR_ptpSEDfit_3DHST}. Bottom panel shows comparison between the $M_{*}$ estimated using our method ($M_{*,\text{ptpSEDfit}}$) and $M_{*}$ taken from the 3D-HST catalog ($M_{*,\text{3D-HST}}$). As shown in the top panel, large discrepancy only seen for a few galaxies and no mixing among the galaxy groups, such that on average $\text{SFR}_{\text{z1-}\Delta \text{MS1}}>\text{SFR}_{\text{z1-}\Delta \text{MS2}}>\text{SFR}_{\text{z1-}\Delta \text{MS3}}$ in both $\text{SFR}_\text{ptpSEDfit}$ and $\text{SFR}_\text{UV+IR}$. A systematic discrepancy is shown in the estimated total $M_{*}$ between the two methods, where the $M_{*,\text{ptpSEDfit}}$ systematically larger than the $M_{*,\text{3D-HST}}$. The larger value of $M_{*}$ from our method moves the sample galaxies rightward on the SFR vs $M_{*}$. Combination of these two discrepancies makes different distribution of the sample galaxies on the SFR versus $M_{*}$ presented in Fig.~\ref{fig:galaxies_sample} and Fig.~\ref{fig:integrated_SFMS}.

\begin{figure}
\centering
\includegraphics[width=0.23\textwidth]{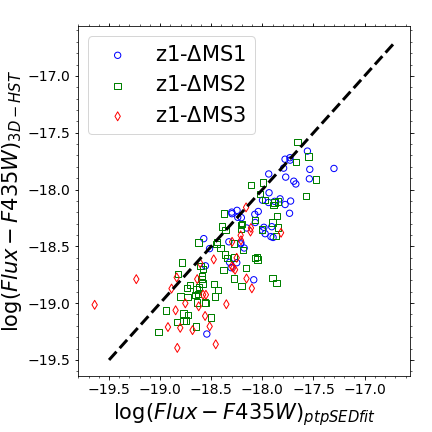}
\includegraphics[width=0.23\textwidth]{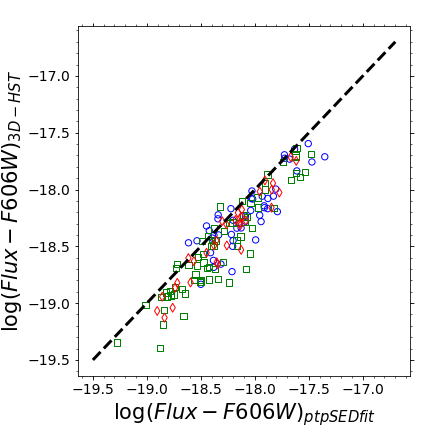}
\includegraphics[width=0.23\textwidth]{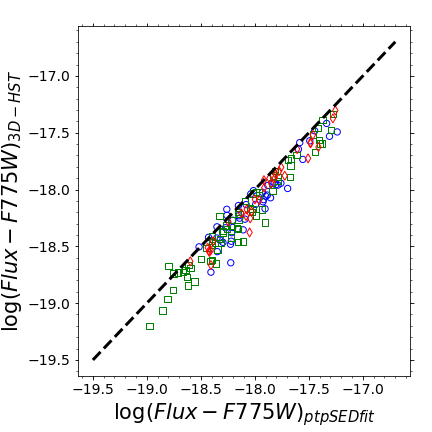}
\includegraphics[width=0.23\textwidth]{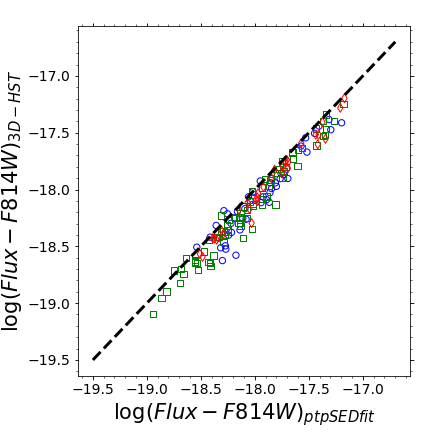}
\includegraphics[width=0.23\textwidth]{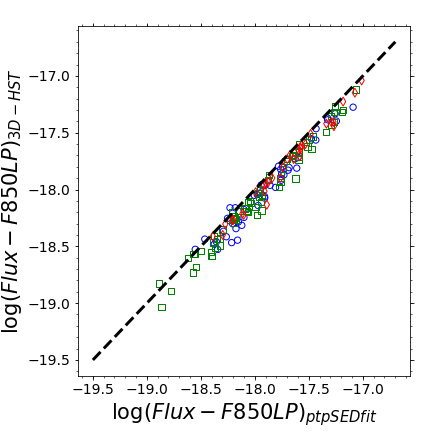}
\includegraphics[width=0.23\textwidth]{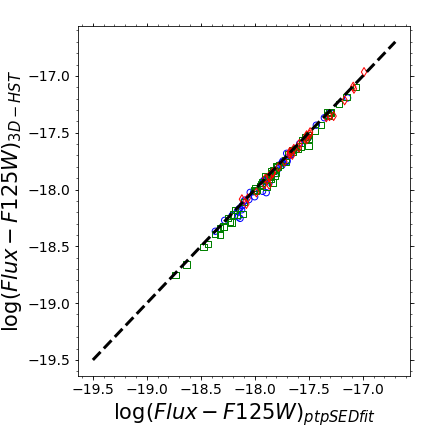}
\includegraphics[width=0.23\textwidth]{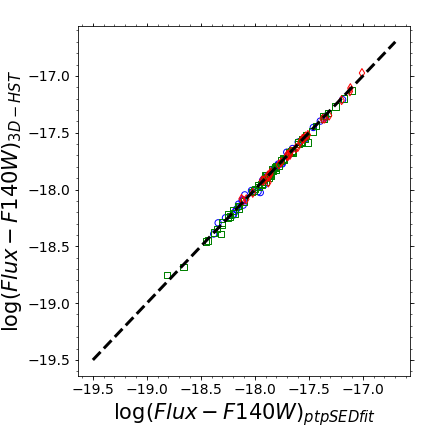}
\includegraphics[width=0.23\textwidth]{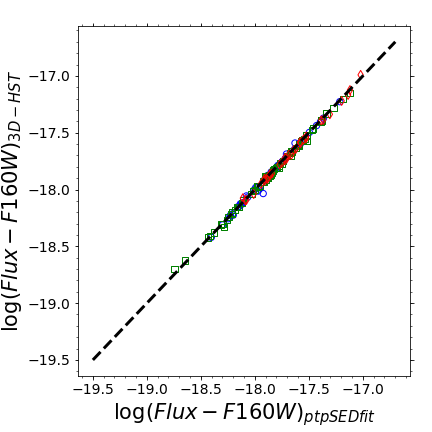}
\caption{Comparisons between integrated fluxes in 8 bands derived by summing up fluxes of galaxy's pixels (this work; x axis in each panel) and those taken from the 3D-HST catalog (which is derived through aperture photometry; y axis in each panel). Blue circles, green squares, and red diamonds represent z1-$\Delta$MS1, z1-$\Delta$MS2, and z1-$\Delta$MS3 sub-samples, respectively. \label{aperture_bias}}
\end{figure}

Part of the discrepancies in $M_{*}$ and SFR are caused by a difference in photometry. Fig.~\ref{aperture_bias} shows comparisons between total fluxes calculated by summing up fluxes of galaxy's pixels (this work) and the integrated fluxes taken from the 3D-HST catalog. As shown in Fig.~\ref{aperture_bias}, there is a discrepancy of total fluxes especially in the bands with shorter wavelength. The 3D-HST photometry is based on $0.7$ arcsec aperture (calculated using  \texttt{SExtractor}) which then extrapolated using surface brightness profile in F160W \citep[see][]{skelton2014}. With this extrapolation, the aperture photometry possibly under estimates the total flux in shorter wavelength band given the extended and clumpy feature of the galaxy structure in the rest-frame UV bands. This discrepancy of total flux in shorter wavelength bands makes discrepancies in color and normalization of the SED and lead to the discrepancy in $M_{*}$ and SFR. 

\begin{figure}
\centering
\includegraphics[width=0.5\textwidth]{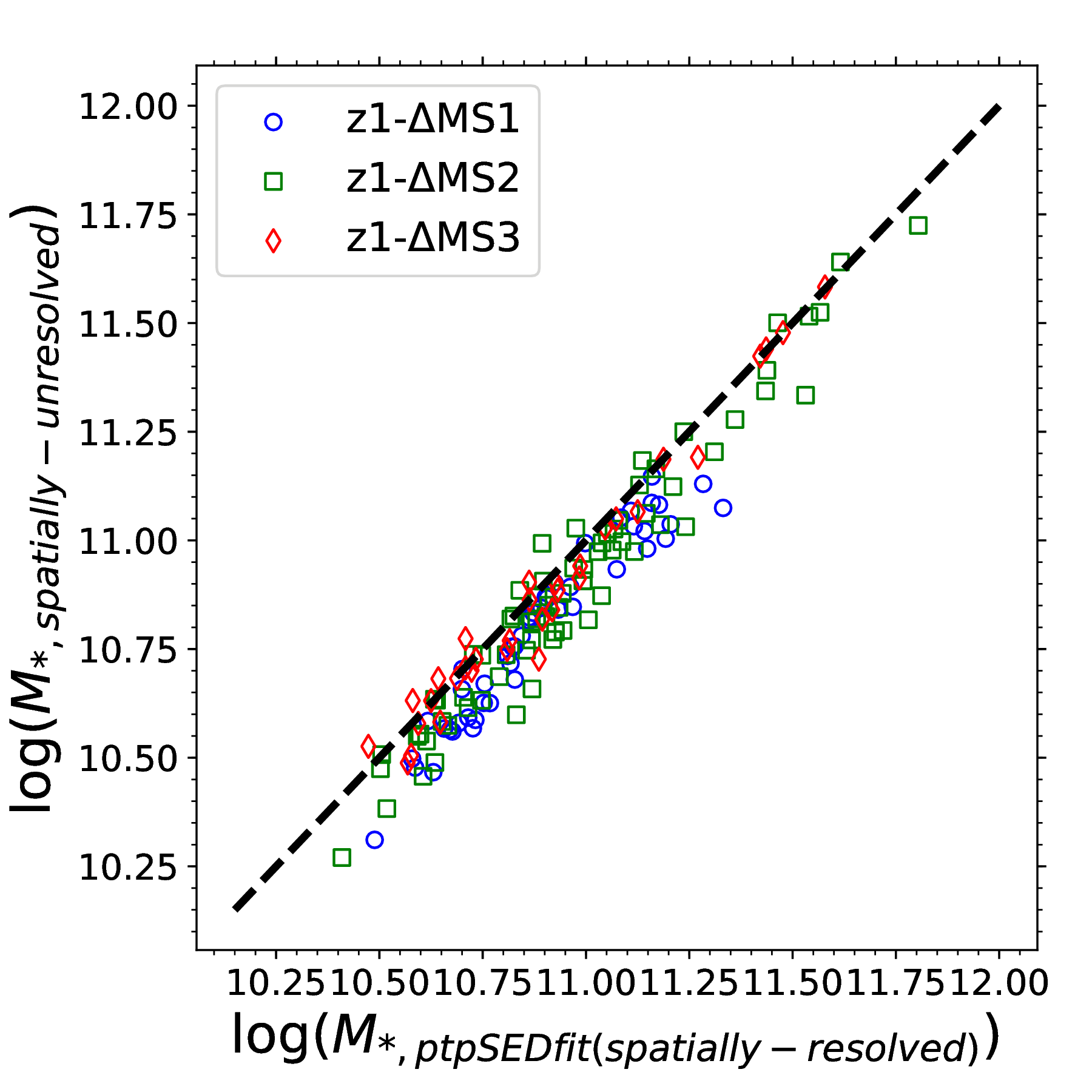}
\caption{Comparison between the integrated $M_{*}$ obtained by summing up the $M_{*}$ of galaxy's pixels that were obtained from the pixel-to-pixel SED fitting ($M_{*,\text{ptpSEDfit(spatially-resolved)}}$) and the $M_{*}$ derived by fitting the integrated SED of the galaxy ($M_{*,\text{spatially-unresolved}}$). Blue circles, green squares, and red diamonds represent z1-$\Delta$MS1, z1-$\Delta$MS2, and z1-$\Delta$MS3 galaxies. \label{compare_SM_ptpSEDfit_spatiallyunresolved} }
\end{figure}

Other possible contributor to the discrepancy in $M_{*}$ is an existence of discrepancy between $M_{*}$ derived from global SED fitting and spatially resolved SED fitting, even if there is no discrepancy in the photometric SED, as observed by previous researchers, e.g. \citet{sorba2015} and \citet{sorba2018}. Fig.~\ref{compare_SM_ptpSEDfit_spatiallyunresolved} shows comparison between the integrated $M_{*}$ derived from the spatially resolved SED fitting ($M_{*,\text{ptpSEDfit(spatially-resolved)}}$, i.e. summing up $M_{*}$ of galaxy's pixels that were derived using the pixel-to-pixel SED fitting) and that derived from a global SED fitting ($M_{*,\text{spatially-unresolved}}$). For the latter, the same fitting method, as the one adopted in the pixel-to-pixel SED fitting method is applied to the integrated SEDs of the sample galaxies. The $M_{*,\text{ptpSEDfit(spatially-resolved)}}$ is systematically higher than the $M_{*,\text{spatially-unresolved}}$.     

\section{Integrated and spatially resolved $UVJ$ diagram}

\begin{figure}
\centering
\includegraphics[width=0.5\textwidth]{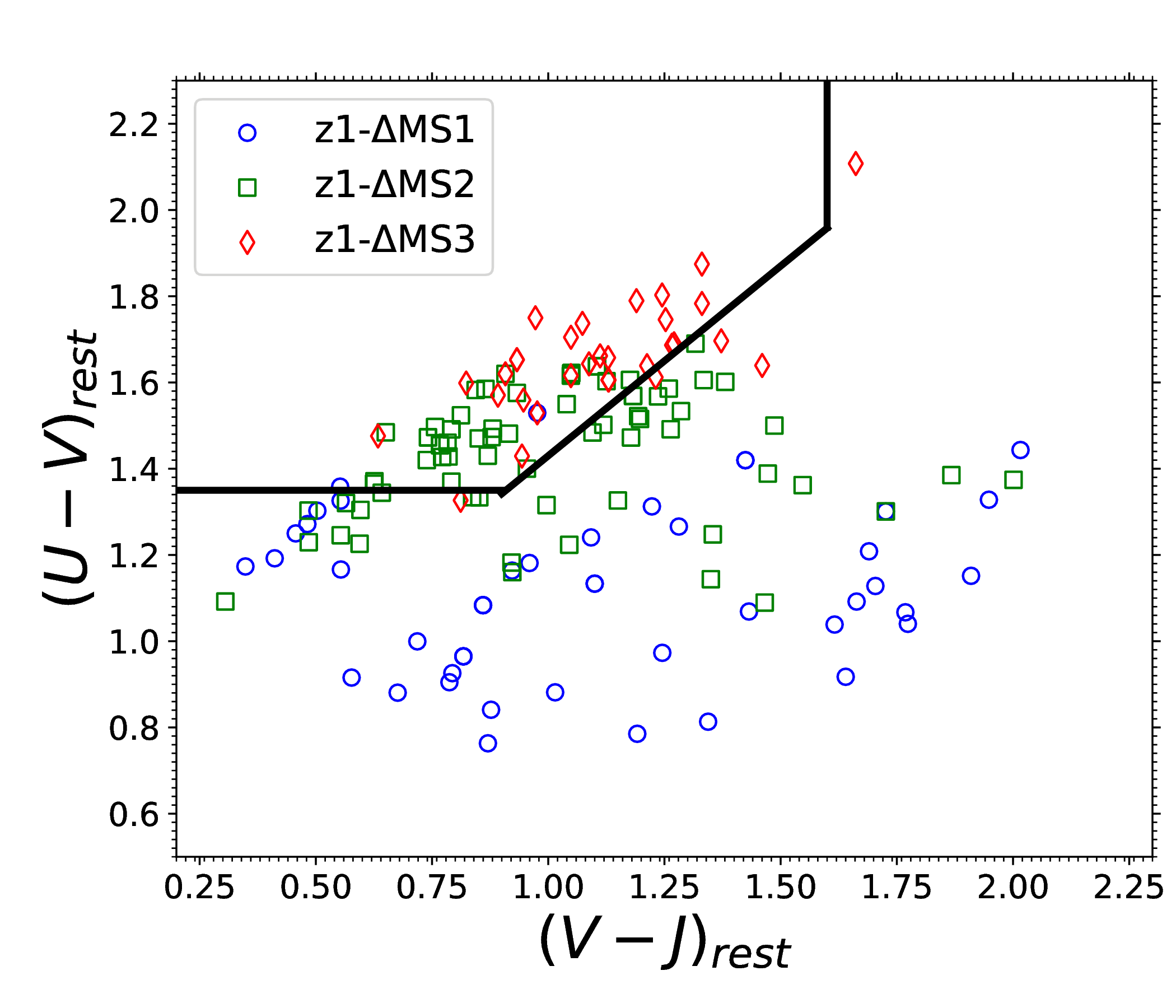}
\caption{Distribution of sample galaxies on the rest-frame $U-V$ versus $V-J$ plane (i.e, $UVJ$ diagram). Upper left "box" represents selection criteria for quiescent galaxies by \citet{williams2009}. Blue circles, green squares, and red diamonds represent z1-$\Delta$MS1, z1-$\Delta$MS2, and z1-$\Delta$MS3 sub-samples, respectively.\label{UVJ_integrated_SED}}
\end{figure}

In order to examine whether quiescent sub-sample galaxies (z1-$\Delta$MS3) are indeed quiescent galaxies and not mistaken for red colours of dusty star-forming galaxies, we estimated the rest-frame $U$, $V$, and $J$ magnitudes of the sample galaxies to check their positions on the $U-V$ versus $V-J$ plane (i.e. $UVJ$ diagram). Top panel of Fig.~\ref{UVJ_integrated_SED} shows positions of the z1-$\Delta$MS1, z1-$\Delta$MS2, and z1-$\Delta$MS3 sub-samples on the $UVJ$ diagram. Upper left region is a selection criteria for quiescent galaxies by \citet{williams2009}. Dusty star-forming galaxies are expected to be in the upper right region of the diagram. We can see that majority of the z1-$\Delta$MS3 galaxies are fall within the selection criteria which confirms that they are indeed quiescent galaxies which dominated by old stellar population. The rest-frame $U$, $V$, and $J$ magnitudes are estimated based on the best-fitting spectrum of the integrated SED (sum of fluxes of galaxy's pixels) obtained from $\chi^{2}$ minimization.

\begin{figure*}
\centering
\includegraphics[width=0.32\textwidth]{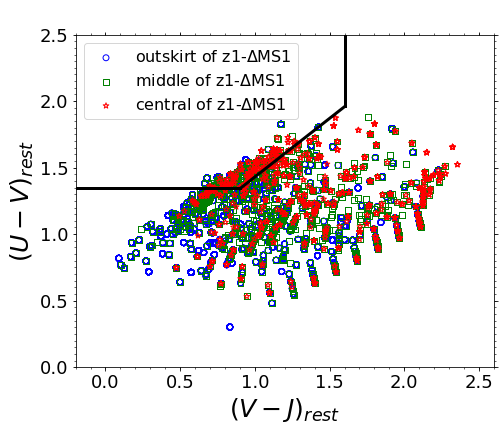}
\includegraphics[width=0.32\textwidth]{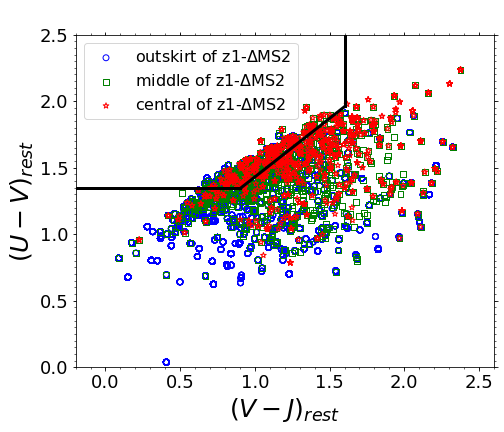}
\includegraphics[width=0.32\textwidth]{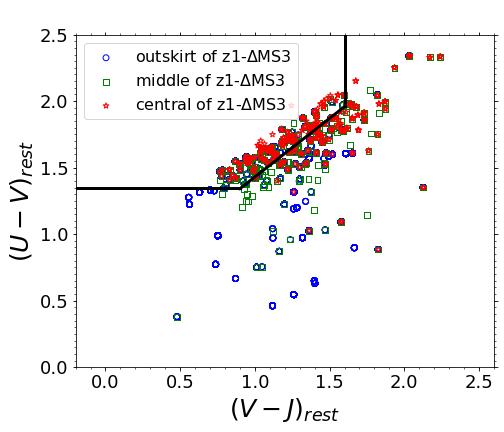}
\caption{Distributons of pixels associated with the z1-$\Delta$MS1 (left panel), z1-$\Delta$MS2 (middle panel), and z1-$\Delta$MS3 (right panel) on the $UVJ$ diagram. Red stars, green squares, and blue circles in each panel represent pixels located in the central ($r\leqslant 2$ kpc), middle ($2 < r \leqslant 8$ kpc), and outskirt ($r>8$ kpc) regions. Upper left "box" in each panel represents selection criteria for quiescent galaxies by \citet{williams2009}. \label{UVJ_inmidout_SEDs}}
\end{figure*}

We also use the above procedure to examine reliability of the centrally quiescent properties of the z1-$\Delta$MS2 and z1-$\Delta$MS3 galaxies as indicated in Fig.~\ref{fig:radial_profile_functionSFMS}. First, we perform SED fitting with $\chi^{2}$ minimization for each SED of the galaxy's bin (collection of pixels) to obtain best-fitting model spectrum of the bin's SED and then calculate $U$, $V$, and $J$ magnitudes based on the best-fitting model spectrum. The magnitudes of a bin are then shared by pixels that belong to the bin, such that all pixels in the bin have the same magnitudes. Fig.~\ref{UVJ_inmidout_SEDs} shows the $UVJ$ diagram. Left panel, middle panel, and right panel show the diagram for the z1-$\Delta$MS1, z1-$\Delta$MS2, and z1-$\Delta$MS3 galaxies, respectively. In each panel, blue circles, green squares, and red stars represent central ($r\leq 2$ kpc), middle ($2 < r \leq 8$ kpc), and outskirt ($r>8$ kpc) regions, respectively. 

Those figures suggest that the central regions of the z1-$\Delta$MS2 and z1-$\Delta$MS3 are less star-forming (i.e. quiescent systems that dominated by old stellar population) compared to their middle and outskirt regions as majority of the central pixels in those galaxies are shifted toward the selection criteria of the quiescent galaxies drawn as a "box" in upper left side in each panel. However, we notice that some central pixels of the z1-$\Delta$MS2 and z1-$\Delta$MS3 galaxies fall into the dusty star-forming locus. Majority of the central pixels of the z1-$\Delta$MS1 are star-forming regions similar as their middle and outskirt pixels, confirming their flat sSFR$(r)$ radial profile. Selection for the radius by which central, middle, and outskirt regions are defined is arbitrary.

%%%%%%%%%%%%%%%%%%%%%%%%%%%%%%%%%%%%%%%%%%%%%%%%%%

% Don't change these lines
\bsp	% typesetting comment
\label{lastpage}
\end{document}